\documentclass[12pt,nofootinbib]{revtex4}
\usepackage{amsmath}
\usepackage{xcolor}
\usepackage{amssymb}
\usepackage{epsfig}
\usepackage{verbatim}
\usepackage{tabularx}
\usepackage{graphicx}
\usepackage{inputenc}
\usepackage{url}
\usepackage{breqn}
\usepackage{hyperref}
\usepackage{footmisc}
\usepackage[page]{appendix}

\def\simlt{\stackrel{<}{{}_\sim}}
\def\simgt{\stackrel{>}{{}_\sim}}

\begin{document}
\begin{titlepage}
\begin{center}
\hfill EFI-15-37
\end{center}
\vspace{1.cm}
\title{Probing the Electroweak Phase Transition at the LHC}
\vspace{1.5cm}

\author{\textbf{Peisi Huang$^{a,c}$, Aniket Joglekar $^{a}$, Bing Li $^{a}$ and Carlos E.M. Wagner$^{a,b,c}$} \\
\vspace{1.5cm}
\normalsize\emph{$^a$Enrico Fermi Institute \& $^b$Kavli Institute for Cosmological Physics,}\\
\normalsize\emph{University of Chicago, Chicago, IL 60637,} \\
\normalsize\emph{$^c$HEP Division, Argonne National Laboratory, 9700 Cass Ave., Argonne, IL 60439.}
\vspace{1.5cm}
}

\begin{abstract}
We study the correlation between the value of the triple Higgs coupling and the nature of the electroweak phase transition. We use an effective potential approach, including higher order, non-renormalizable terms coming from integrating out new physics. We show that if only the dimension six operators are considered, large positive deviations of the triple Higgs coupling from its Standard Model (SM) value are predicted in the regions of parameter space consistent with a strong first order electroweak phase transition (SFOEPT). We also show that at higher orders sizable and negative deviations of the triple Higgs coupling may be obtained, and the sign of the corrections tends to be correlated with the order of the phase transition. We also consider a singlet extension of the SM, which allows us to
establish the connection with the effective field theory (EFT) approach and analyze the limits of its validity.  Furthermore, we study how to probe the triple Higgs coupling from the double Higgs production at the LHC.  We show that selective cuts in the invariant mass of the
two Higgs bosons should be used, to maximize the sensitivity for values of the triple Higgs
coupling significantly different from the Standard Model one.
\end{abstract}
\maketitle
\end{titlepage}
\section{Introduction}
After the Higgs discovery at the LHC~\cite{ATLASHiggs, CMSHiggs}, the Higgs properties, including the Higgs mass and the Higgs couplings to the Standard Model (SM) particles have been measured~\cite{HiggsMass,ATLAScouplings,CMScouplings}. Those measurements show that the Higgs boson properties are close to the SM ones. Those properties are related to the gauge transformation properties of the Higgs field and with the mechanism of electroweak symmetry breaking, but provide little information about the properties of the Higgs potential. In the SM, a quadratic coupling and a quartic coupling completely specify this potential. In the theories beyond the SM, there can be contributions to the effective potential from the higher dimensional operators, with an effective cut-off given by the characteristic new physics scale of the theory. As a result, the self-interactions of the Higgs field, most notably the triple Higgs coupling $\left(\lambda_3\right)$, are modified. 

What makes the deviation of $\lambda_3$ from its SM value even more exciting is that $\lambda_3$ is closely related to the strength of the electroweak phase transition (EPT)~\cite{Menon:2004wv,Grojean:2004xa,Noble:2007kk,Barger:2011vm,Chung:2012vg,Katz:2014bha,Curtin:2014jma,He:2015spf}. Understanding the nature of the EPT will advance our knowledge of the possible realization of electroweak baryogenesis\cite{Sakharov:1967dj}, which is an attractive explanation of the baryon anti-baryon asymmetry, that can only happen if the EPT is first order. Today the electroweak symmetry is clearly broken, while in the early universe the $SU(2)\times U(1)$ symmetry was preserved, a result that may be easily understood considering the finite temperature effects to the effective potential. About $\textrm{10}^{\textrm{-10}}$ seconds after the Big Bang, the universe underwent a phase transition from the unbroken phase to the broken phase. This leads to formation and expansion of  bubbles of the true vacuum configuration in the false, gauge symmetric vacuum. In the presence of CP-violation, particle interactions with the expanding bubbles may lead to the creation of an excess of baryons inside the bubbles by means of baryon number violating processes induced by sphalerons~\cite{Klinkhamer:1984di}. These sphaleron processes, if  they were in equilibrium inside the bubbles, would wipe-off the created excess of baryons. The rate of these processes depend exponentially on the ratio of the vacuum expectation value (VEV) to the critical temperature at the time of the phase transition and are suppressed if the phase transition is of strongly first order~\cite{Shaposhnikov:1987tw}. Unfortunately, in the pure SM scenario, the requirement of a sufficiently strong first order phase transition translates into an upper bound on the Higgs mass of about 35 GeV~\cite{Dine:1992vs,Kajantie:1995kf}. The discovery of the Higgs boson at 125 GeV excludes such a simple scenario~\cite{Chatrchyan:2012xdj,Aad:2012tfa}. This motivates a further investigation of the viability of the electroweak baryogenesis in minimally extended scenarios. 

 A first order electroweak phase transition (FOEPT) may lead to the production of gravitational waves, but the characteristic scales associated with it make their detection very difficult, albeit not impossible, to detect in the near future~\cite{Apreda:2001us,Grojean:2006bp,Leitao:2012tx,Kikuta:2014eja,Kakizaki:2015wua}.  Alternatively, the models that lead to a FOEPT through a relevant modification of the zero temperature effective potential  can be probed from the deviation of $\lambda_3$ from its SM value, as suggested in  previous studies~\cite{Noble:2007kk,Katz:2014bha,Curtin:2014jma}.

At the LHC, $\lambda_3$ can be probed by the process of double Higgs production. Mainly due to the destructive interference between the one-loop diagrams, the production cross section reduces initially, as the $\lambda_3$ is enhanced from its SM value. At the next-to-leading order, the minimum occurs for $\lambda_3\sim2.45\,\lambda_3^{SM}$~\cite{Frederix:2014hta}. Further enhancement of the~$\lambda_3$ value increases the cross section again, which exceeds the SM value for~$\lambda_3>5\lambda_3^{SM}$. The cross-section also increases if the correction to $\lambda_3^{SM}$ is negative. The $b\bar{b}\gamma\gamma$, $b\bar{b}\tau^+\tau^-$, $b\bar{b}W^+W^-$ and $bb\bar{b}\bar{b}$ channels~\cite{Baur:2003gp,Dolan:2012rv,Baglio:2012np,Dolan:2012ac,Barger:2013jfa,Barr:2013tda,Yao:2013ika,ATL-PHYS-PUB-2014-019,Lu:2015jza} have been studied. These studies showed that around 50$\%$ accuracy can be achieved from the $b\bar{b}\gamma\gamma$ channel alone assuming that $\lambda_3$ is not too far away from its SM value and the acceptance for different values of $\lambda_3$ stays the same. However, as pointed out in~\cite{Barger:2013jfa}, the acceptance drops significantly for large values of $\lambda_3$. In this article we perform a detailed study of the impact of a large deviation from $\lambda_3^{SM}$ on the double Higgs production process. We also present an analysis of the LHC searches for this process including relevant QCD background contributions that have been overlooked in the previous studies.

The organization of this article is as follows : In Sec.~\ref{sec:EWPT}, we calculate the values of $\lambda_3$ if the EPT is first order in a simplified model, where we include higher order terms in the effective potential. In Sec.~\ref{sec:singlettotal}, we compare our results to those obtained in singlet extensions like the ones that may be obtained from the scalar Higgs sector in the Next to Minimal Supersymmetric Standard Model (NMSSM). In Sec.~\ref{sec:LHC}, we discuss the measurement of $\lambda_3$ at the LHC, for the SM-like values as well as for values of $\lambda_3$ that present a large positive or negative deviation with respect to the SM value. We reserve Sec.~\ref{conclusions} for the conclusions and some technical details to the Appendices. 

\section{The Effective  Potential and the Trilinear Higgs Coupling}
\label{sec:EWPT}


A modification of the nature of the phase transition may be achieved by adding extra terms to the Higgs potential~\cite{Cohen:1993nk,Riotto:1999yt,Morrissey:2012db}. These may appear through relevant temperature dependent modifications of the Higgs potential, beyond those associated with the increase of the effective mass parameter, which lead to the symmetry restoration phenomenon (see, for example, Refs.~\cite{ Cline:1995dg, Carena:1996wj,deCarlos:1997ru,Carena:1997ki,Laine:1998qk,Carena:2008vj, Cohen:2011ap,Laine:2012jy,Curtin:2012aa,Carena:2012np,Carena:2004ha,Huang:2012wn,
Fok:2008yg,Davoudiasl:2012tu}).  

Alternatively, these effects may be already present at zero temperature, through additional terms in the Higgs potential induced by integrating out new physics at the scales above the weak scale. In this section we concentrate on the second possibility and illustrate the impact of such additional terms on the enhancement of $\lambda_3$ in minimally extended  models. 
Several simple extensions of the SM are capable of generating the required extra terms in the potential and have been studied in the literature~\cite{Pietroni:1992in,Davies:1996qn,Barger:2011vm,Huber:2006wf,Carena:2011jy,
Chung:2012vg,Huang:2014ifa,Noble:2007kk,Katz:2014bha,Curtin:2014jma,He:2015spf,Menon:2004wv,
Grojean:2004xa}. In Sec.~\ref{sec:singlettotal}, we analyze one such example, where a gauge singlet is added to the SM. This can lead to a relevant modification of the trilinear Higgs coupling  with respect to the SM value $\lambda_3^{SM}$, even for values of the singlet mass much larger than the weak scale. In such a case, the singlet decouples from physics processes at the LHC, allowing a comparison of these results with the ones obtained in the effective low energy field theory.

In this section, we take a general approach to the effective field theory (EFT), where non-renormalizable terms are added to the Higgs potential. We investigate whether these can potentially generate considerably larger cross-sections for $gg\rightarrow hh$ process compared to the standard model.  We also explore the possibility of these being compatible with a strongly first order electroweak phase transition (SFOEPT). Such modifications to~$\lambda_3^{SM}$  would make for a viable probe to the new physics at the LHC and beyond. 

\subsection{Non-renormalizable terms in the low energy Higgs potential}
\label{subsec:toy}
The general formalism in this section is as follows. All the tree-level effective operators represented by powers of $\left(\phi^\dagger\phi\right)$ are added to the usual Higgs potential at the temperature $T=0$ as follows
\begin{align}\label{eqn:GenPot}
V\left(\phi,0\right)=& \frac{m^2}{2}(\phi^\dagger\phi) + \frac{\lambda}{4}(\phi^\dagger\phi)^4 + \sum^\infty_{n=1}\frac{c_{2n+4}}{2^{(n+2)}\Lambda^{2n}}\left(\phi^\dagger\phi\right)^{n+2},\end{align}
where $\phi = v + h$ and hence the VEV is given as $\left<\phi\right> = 246$~GeV.
This leads to a correction to the SM value of the triple Higgs coupling as shown in the Appendix~\ref{app:A}.
\begin{align}
\lambda_3=&\frac{3m^2_h}{v}\left(1+\frac{8v^2}{3m_h^2}\sum^\infty_{n=1}\frac{n(n+1)(n+2)c_{2n+4}v^{2n}}{2^{n+2}\Lambda^{2n}}\right).\label{eqn:triple}
\end{align}

The non-zero temperature effects are approximately accounted for by adding a thermal mass correction term to the Higgs potential. This term is generated in the high-T expansion of the one loop thermal potential. At temperature T, we get $m^2(T) = m^2 +a_0 T^2$. We have ignored the small cubic term contributions as well as the logarithmic contributions as they are suppressed compared to the contributions from higher order terms. Here we have assumed that the heavy new physics is not present in the EFT at the weak scale and therefore its contribution is Boltzmann suppressed at the EPT scale. In such a case $a_0$ is a constant proportional to the square of SM gauge and Yukawa coupling constants. Assuming all $c_{2n} \simeq 1$, the minimum value that $\Lambda$ can achieve is $174$ GeV in this formulation, at which point the convergence of the series is lost for values of $\phi$ close to its VEV. However, in any consistent EFT, the cut-off scale $\Lambda$ will be considerably higher than $174$ GeV.

Using Eq.~(\ref{eqn:triple}), we define another quantity~$\delta$ which quantifies the deviations of the trilinear Higgs coupling with respect to the SM value as
\begin{align}
\delta&=\frac{\lambda_3}{\lambda_3^{SM}}-1=\frac{8v^2}{3m_h^2}\sum^\infty_{n=1}\frac{n(n+1)(n+2)c_{2n+4}v^{2n}}{2^{n+2}\Lambda^{2n}},\label{eqn:delta}
\end{align}
where we restrict $|c_{2n+4}|<1$.

The values of the enhancement of $\lambda_3$ for a given $\Lambda$ for all potentials satisfying these conditions are shown in Fig.~\ref{fig:convergence}. This maximal possible value, shown in the the upper-most black (dashed) line in all the panels in Fig.~\ref{fig:convergence}, is obtained assuming all $c_{2n}=1$ and leads to a large enhancement even at a relatively large value of $\Lambda$. However, the only condition that we have imposed on the potential so far is the existence of a local minimum with a second derivative consistent with the measured Higgs mass $m_h \simeq 125$~GeV. For this minimum to represent the physical vacuum of the theory, however,  it should be a global one. As we shall show, the global minimum requirement imposes strong constraints on the possible enhancement of the triple Higgs coupling. 

In our further analysis, we choose not to consider the terms of the order higher than $\left(\phi^\dagger\phi\right)^5$ as they introduce negligible corrections for the cut-offs higher than $v$ as shown in Fig.~\ref{fig:convergence}. We separately analyze the nature of the phase transition and the maximum positive and negative values for $\delta$ in each of the three cases corresponding to $\left(\phi^\dagger\phi\right)^3$, $\left(\phi^\dagger\phi\right)^4$ and $\left(\phi^\dagger\phi\right)^5$.  Let us stress that
these momentum independent operators preserve the custodial symmetry and evade the tight phenomenological constraints coming from the $\rho$ parameter. The momentum dependent non-renormalizable operators~\cite{Corbett:2012ja,Goertz:2014qta,Azatov:2015oxa,He:2015spf}, instead, may contribute to the oblique corrections and are very tightly constrained by the electroweak precision measurements. 
A particularly relevant one for our analysis is
\begin{equation}
\frac{c_H}{8\Lambda^2}  \partial_{\mu} (\phi^{\dagger} \phi)  \partial^{\mu} (\phi^\dagger \phi), 
\end{equation}
This correction plays a relevant role in the singlet case that we shall discuss below, but is also restricted  by Higgs precision measurements and tend to be small. Hence, in most of our analysis we shall ignore the momentum dependent corrections but we
shall consider them in the comparison with the singlet case in section~\ref{sec:singletEFT}.

\subsubsection{Higgs Potential of order $\left(\phi^\dagger\phi\right)^3$}
\label{sec:phi6}

From Eq.~(\ref{eqn:GenPot}) and Eq.~(\ref{eqn:triple}), the potential and the triple Higgs coupling are given by
\begin{align}
V(\phi,T)&=\frac{m^2+a_0T^2}{2}\left(\phi^\dagger\phi\right)+\frac{\lambda}{4}\left(\phi^\dagger\phi\right)^2+\frac{c_6}{8\Lambda^2}\left(\phi^\dagger\phi\right)^3\label{eqn:temp6}\\
\lambda_3&=\frac{3m_h^2}{v}\left(1+\frac{2c_6v^4}{m_h^2\Lambda^2}\right)\label{eqn:enh6}
\end{align}
This case has been studied in the literature in various contexts~\cite{Barger:2011vm, Chung:2012vg,Noble:2007kk,Katz:2014bha,Curtin:2014jma,He:2015spf,Menon:2004wv,Grojean:2004xa}. We point out a few key things pertaining to this case in the present context. 

We require $c_6>0$ for the stability of the potential~\footnote{We understand that even for $c_6 < $ 0 the stability could be recovered for field values that are above the cutoff,
where the EFT is not valid. We will consider the case of $c_6 <$ 0 when we study the $(\phi^\dagger \phi )^{4,5}$ extensions.}. 
The requirement that there should be a minimum of the potential at $\phi=\phi_c$ degenerate with the extreme at $\phi = 0$ for the temperature $T=T_c$ leads to 
\begin{align}
 \lambda^2=4m^2(T_c)\frac{c_6}{\Lambda^2}.\label{eqn:root1}
\end{align} 
This implies that $m^2(T)$, which is the curvature of the potential at $\phi = 0$, should be greater than zero at $T=T_c$ for the phase transition to be of the first order. The minimum of the potential at the critical temperature is at 
\begin{align}
\left(\phi_c^\dagger\phi_c\right)=v_c^2=-\frac{\lambda \Lambda^2}{c_6}.\label{eqn:root2}
\end{align} 
what implies that an additional condition  to obtain a FOEPT is that the effective quartic coupling should be negative, namely $\lambda < 0$. 

The value of the Higgs mass imposes a relation between $\lambda$ and $c_6$, namely
\begin{align}
\lambda+\frac{3 c_6}{2\Lambda^2}v^2&=\frac{m_h^2}{2v^2}\label{eqn:curvature}
\end{align}
Using Eq.~(\ref{eqn:root2}) and Eq.~(\ref{eqn:curvature}) gives
\begin{align}
\frac{c_6}{\Lambda^2}=\frac{m_h^2}{3v^2\left(v^2-\frac{2}{3}v_c^2\right)}\label{eqn:kappa}
\end{align}
From where all coefficients $m^2$, $\lambda$ and $c_6$ may be written in terms of the $m_h$, $v_c$ and $v$. Using these
relations one obtains 
%
\begin{align}
T_c^2=\frac{3 c_6}{4 \Lambda^2 a_0}&\left(v^2-v_c^2\right)\left(v^2-\frac{v_c^2}{3}\right).\label{eqn:critical1}
\end{align}
Demanding both $c_6$ and $T_c^2$ to be positive, we get $v_c<v$. This translates into an upper bound on $c_6$ using Eq.~(\ref{eqn:kappa})
\begin{align}
\frac{c_6}{\Lambda^2}<\frac{m_h^2}{v^4} \ .
\label{eqn:kmax}
\end{align}

Then from the Eq.~(\ref{eqn:enh6}), we conclude that the coupling can be enhanced by a factor of three at most.  
Moreover, demanding $v_c^2 > 0$, or equivalently $\lambda < 0$, puts an additional constraint on the obtention of a FOEPT, namely 
\begin{align}
\frac{c_6}{\Lambda^2}  >  \frac{m_h^2}{3 v^4}
\label{eon:c6min}
\end{align}
what implies a minimal enhancement of a factor two thirds.  

This implies that a FOEPT may only be obtained if the following conditions are fulfilled.
\begin{eqnarray}
\label{eq:enhancement}
\frac{2}{3}   \leq  &  \delta &  \leq  2 .
\end{eqnarray}
Moreover, for $c_6=1$, Eq~(\ref{eqn:kmax}) and Eq~(\ref{eon:c6min}) imply a bound on the effective cutoff $\Lambda$, namely
\begin{eqnarray}
\frac{v^2}{m_h}  < &  \Lambda &  < \frac{\sqrt{3}v^2}{m_h}  ,
\label{eq:enhancement2}
\end{eqnarray}
which correspond to upper and lower bounds on $\Lambda$ of approximately 484 GeV and 838 GeV respectively, and larger enhancement $\delta$ is obtained for the smaller values of the cutoff. The phase transition becomes stronger first order for smaller values of the cutoff
and becomes a weakly first order one for values of $\Lambda$ close to the upper 
bound in Eq.~\ref{eq:enhancement2}.  Let us stress that for values of $\Lambda$ below the
lower bound in Eq.~\ref{eq:enhancement2},  $\Lambda < 484$~GeV, the minimum at $T = 0$ is no longer a global minimum and hence  electroweak symmetry breaking does not occur. 

\begin{figure}[ht]
\includegraphics[width = 7.5cm, clip]{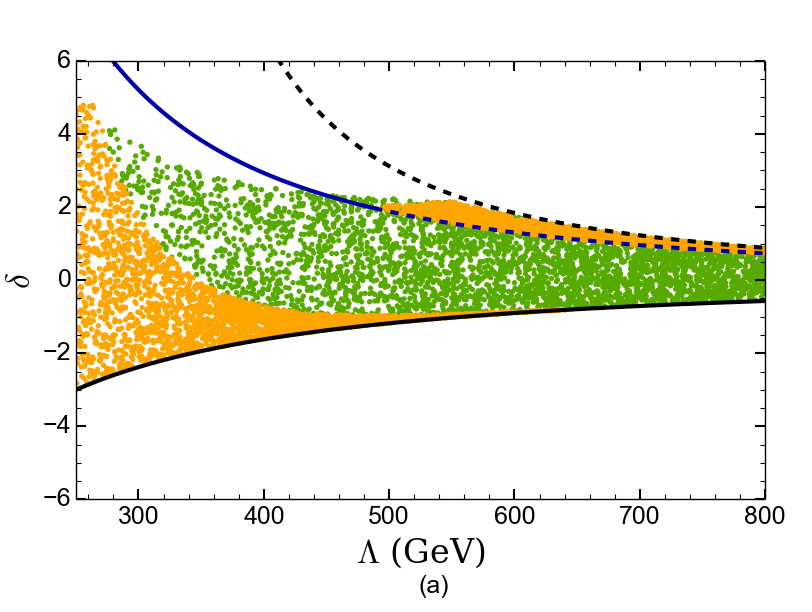}
\includegraphics[width = 7.5cm, clip]{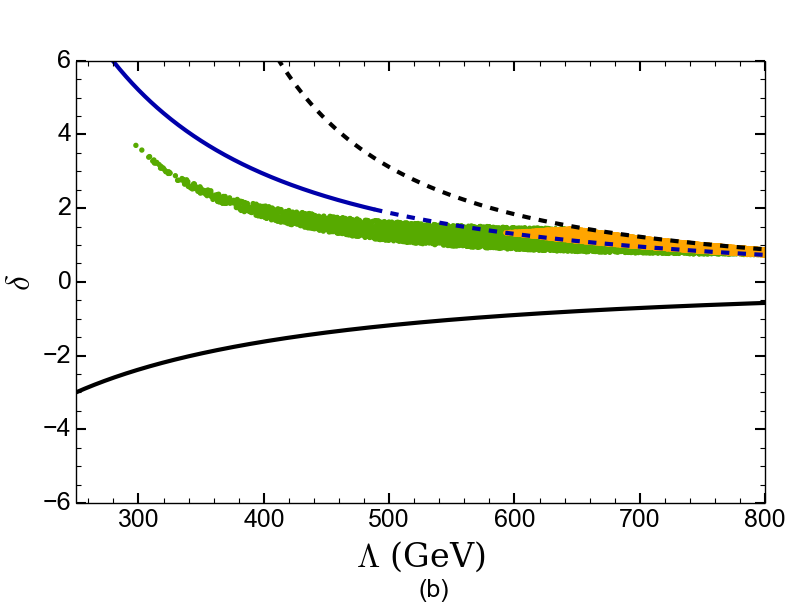}\\
\includegraphics[width = 7.5cm, clip]{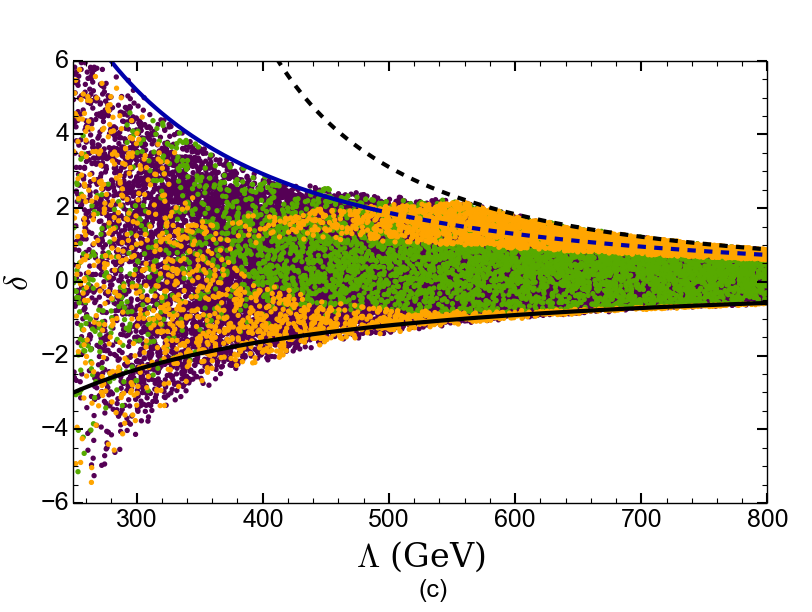}
\includegraphics[width = 7.5cm, clip]{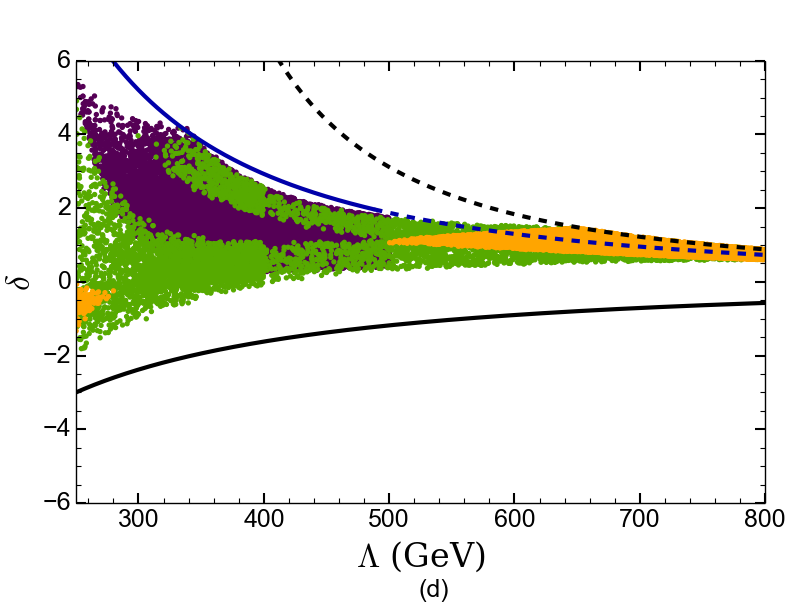}
\caption{Triple Higgs coupling correction $\delta$ as a function of the cutoff $\Lambda$. The upper dashed black line shows the maximum value of $\delta$ for the infinite sum with all $|c_{2n}|=1$. The dashed dark blue shows the values consistent with a FOEPT for the $\left(\phi^\dagger\phi\right)^3$ potential extension, for $c_6 =1$, while for the same conditions solid light blue line is forbidden due to the absence of electroweak symmetry breakdown. Fig.~1(a) and 1(b) show the results  for the $\left(\phi^\dagger\phi\right)^4$ potential. The different colors correspond to the different hierarchies of the effective potential coefficients as explained in the text. Fig.1(a) shows the general case while the Fig.~1(b) shows the result if a first order electroweak phase transition (FOEPT) is demanded. Fig.~1(c) and 1(d) show similar results but for the  $\left(\phi^\dagger\phi\right)^5$ potential, with different colors again corresponding to different coefficient hierarchies defined in the text.  The lower solid black line shows the maximal negative values of $\delta$ possible for the order $\left(\phi^\dagger\phi\right)^4$ potential.}
\label{fig:convergence}
\end{figure}

In Fig~\ref{fig:convergence}, we show the possible triple Higgs coupling enhancement factor $\delta$ as a function of the cutoff $\Lambda$ for different extensions of the SM effective potential. The particular case of the potential of order $\left(\phi^\dagger\phi\right)^3$ is represented by the blue curve. The maximum enhancement $\lambda_3=3\lambda_3^{SM}$ is achieved at $\Lambda\sim484$ GeV. For the cut-offs above $\Lambda\sim 838$~GeV, not shown in the figure, the phase transition is not first order anymore, but the Higgs potential is still a viable one. Note that the low value of the cut-off does not necessarily correspond to any physical mass scale, as will be discussed in the singlet case, in Sec.~\ref{sec:singlettotal}. 


Let us note before closing that 
in Ref.~\cite{Gupta:2013zza} it is found that  for a FOEPT to take place, the enhancement due to a six-dimensional operator to the Higgs potential cannot be larger than $\sim20\%$.  In order to understand the difference of their result with ours we notice that in their normalization, the coefficient of the $(\phi^\dagger\phi)^3$ term is written as $\frac{\bar{c}_6\lambda}{f^2}$, where $\lambda$ is the coefficient of the $(\phi^\dagger\phi)^2$ term\footnote{We denote the coefficient used in Ref.~\cite{Gupta:2013zza}   $\bar{c}_6$, not to confuse it with the coefficient $c_6$ defined above.}. The discrepancy is due to  the assumption in Ref.~\cite{Gupta:2013zza} that $\bar{c}_6>0$ and $\bar{c}_6 v^2/f^2$ small.  As we showed above, for a FOEPT to take place, the effective quartic coupling~$\lambda<0$, which means $\bar{c}_6<0$ is required for the stability of the potential. Also, for $\lambda < 0$, the required condition to obtain a positive Higgs mass is $\bar{c}_6 v^2/f^2<-\frac{2}{3}$.  
Thus, in the notation of Ref.~\cite{Gupta:2013zza}, $|\bar{c}_6| v^2/f^2$ cannot be used as a small expansion parameter in the region of parameters consistent with a FOEPT.   Finally, the
upper bound assumed on $\bar{c}_6/\Lambda^2$, coming from Ref.~\cite{Grojean:2004xa},  is similar to the one we derived in Eq.~(\ref{eqn:kmax}) and is applicable to $c_6/\Lambda^2$ and not to $\bar{c}_6/\Lambda^2$.

\subsubsection{Higgs Potential of order $\left(\phi^\dagger\phi\right)^4$}

From Eq.~(\ref{eqn:GenPot}) and Eq.~(\ref{eqn:triple}), the potential and the triple Higgs coupling are
\begin{align}
V(\phi,T)&=\frac{m^2+a_0T^2}{2}\left(\phi^\dagger\phi\right)+\frac{\lambda}{4}\left(\phi^\dagger\phi\right)^2+\frac{c_6}{8\Lambda^2}\left(\phi^\dagger\phi\right)^3+\frac{c_8}{16\Lambda^2}\left(\phi^\dagger\phi\right)^4\label{eqn:pot8}\\
\lambda_3&=\frac{3m_h^2}{v}\left(1+\frac{2c_6v^4}{m_h^2\Lambda^2}+\frac{4c_8v^6}{m_h^2\Lambda^4}\right)\label{eqn:enh8}
\end{align}

This case is particularly interesting because contrary to the $(\phi^{\dagger} \phi)^3$ case,
the trilinear Higgs couplings may be either enhanced or suppressed and one can even get
an inversion of the sign of $\lambda_3$ with respect to $\lambda_3^{SM}$.  As mentioned before, a suppression or change of sign of $\lambda_3$ would be interesting from the collider perspective as it avoids the problem of a strong destructive interference between the box and the triangle diagrams for $gg\rightarrow hh$. 

The orange and green regions in Fig.~\ref{fig:convergence}(a) and Fig.~\ref{fig:convergence}(b) correspond to the regions consistent with the experimental values of the Higgs mass and the Higgs VEV. Fig.~\ref{fig:convergence}(a) shows the possible modifications ($\delta$) of the $\lambda_3^{SM}$ possible in this case. Fig.~\ref{fig:convergence}(b) outlines the region in Fig.~\ref{fig:convergence}(a) which corresponds to the FOEPT. This shows that an inversion of sign or suppression of $\lambda_3$ with respect to $\lambda_3^{SM}$ necessarily implies that the phase transition is not a first order one. In the construction of Fig.~\ref{fig:convergence}(b), we have not considered the region of the parameter space corresponding to potentials with barriers between the minima at $\phi=0$ and $\phi=v$ at $T=0$. This is due to the fact that a metastability analysis would be required to determine the part of this region in which a FOEPT takes place. Therefore, this rather small region is neglected in our analysis. As a result of this, a small part of the dashed blue curve is not surrounded by the shaded regions. The same is true for Fig.~\ref{fig:convergence}(d).

In Fig.~\ref{fig:convergence}(a) and Fig.~\ref{fig:convergence}(b), the different colors indicate different regions of the parameter space. The orange region corresponds to $|c_6|=1,\,0<c_8<1$, while the green region corresponds to $|c_6|<1,\,c_8=1$. The regions can overlap, because a different combination of $c_6$ and $c_8$ can produce the same value of $\delta$ for the same cut-off. In fact, beneath all of the orange region above the blue curve, there exists a green region. We observe that it is possible to obtain $\lambda_3$ values ranging from $-2\lambda_3^{SM}$ to $6\lambda_3^{SM}$ for cut-offs higher than $250$ GeV. Demanding a FOEPT reduces it to a smaller  range from $\frac{5}{3}\lambda_3^{SM}$ to~$5\lambda_3^{SM}$. We also note from Fig.~\ref{fig:convergence}(b) that the FOEPT has a lower bound on the cut-off $\sim 300$ GeV, which is somewhat lower than in the $(\phi^\dagger \phi)^3$ case. Note that, the contribution to $\lambda_3$ from the dim-8 operators is suppressed compared to that from the dim-6 operators. The fact that in a $(\phi^\dagger\phi)^4$ theory, $\lambda_3$ has a much larger range in the general case compared to a $(\phi^\dagger\phi)^3$ theory, and in the region consistent with the FOEPT is because with $c_8$ being a positive number, $c_6$ is allowed to take negative values in the range of $|c_6|<1$ in a $(\phi^\dagger\phi)^4$ theory, while $0<c_6<1$ has to be fulfilled in a $(\phi^\dagger\phi)^3$ theory. 

Let us stress that negative values of $\delta$ imply that the curvature is decreasing at $\phi=v$. If this behavior is preserved at larger values of $\phi$, one would expect a maximum of the potential for $\phi>v$. Then the stability of the potential means there has to be one more minimum for $\phi>v$. The deeper the extra minimum, the more negative is the value of $\lambda_3$. Thus, demanding the physical minimum to be a global one, a maximal negative value would occur at the point where both minima have the same potential value. 
In order to retain the analytic control, we plot the analytical bound coming from the  point marking the end of the absolute stability. For $(\phi^\dagger\phi)^4$ case, this bound is the black curve at the bottom of each panel of Fig.~\ref{fig:convergence}. As shown in appendix~\ref{app:C}, this maximally negative enhancement is given as
\begin{align}
\delta > -\frac{x}{1+\sqrt{1+x}},\quad\text{where}\quad x=\frac{4v^4}{m_h^2\Lambda^2} .
\label{eqn:lwrbnd}
\end{align} 
Observe, however, that for $\Lambda \simeq 250$~GeV the second minimum would occur at values of $\phi$ of order or larger than $\Lambda$, and hence this analytical result should be taken with care.  The numerical results of Fig.~\ref{fig:convergence} were obtained by only demanding the physical minimum to be the global one. The largest negative enhancements obtained numerically are consistent with the predictions of  Eq.~(\ref{eqn:lwrbnd}) up to values of $\Lambda \simeq v$.  Let us stress again that although we show examples with very low cutoff values,  those low cutoff values may be hard to realize in any realistic model. 


\subsubsection{Higgs Potential of order $\left(\phi^\dagger\phi\right)^5$}

From Eq.~(\ref{eqn:GenPot}) and Eq.~(\ref{eqn:triple}), the potential and the triple Higgs coupling in this case are
\begin{align}
V(\phi,T)=\frac{m^2+a_0T^2}{2}\left(\phi^\dagger\phi\right)&+\frac{\lambda}{4}\left(\phi^\dagger\phi\right)^2+\frac{c_6}{8\Lambda^2}\left(\phi^\dagger\phi\right)^3+\frac{c_8}{16\Lambda^4}\left(\phi^\dagger\phi\right)^4+\frac{c_{10}}{32\Lambda^6}\left(\phi^\dagger\phi\right)^5\\ \notag\\
\lambda_3&=\frac{3m_h^2}{v}\left(1+\frac{2\,c_6v^4}{m_h^2\Lambda^2}+\frac{4\, c_8v^6}{m_h^2\Lambda^4}+\frac{5\,c_{10}v^8}{m_h^2\Lambda^6}\right)\label{eqn:enh10}
\end{align}

Most of the analysis is the same as that for the $\left(\phi^\dagger\phi\right)^4$ case, and the extra minimum develops for $\phi>v$, when the correction to $\lambda_3^{SM}$ is negative. Barring the possibility of metastability, the bound on the maximal negative correction corresponds to the point in which the extra 
minimum is degenerate with the physical one. 

\begin{figure}
\includegraphics[width = 8.0cm, clip]{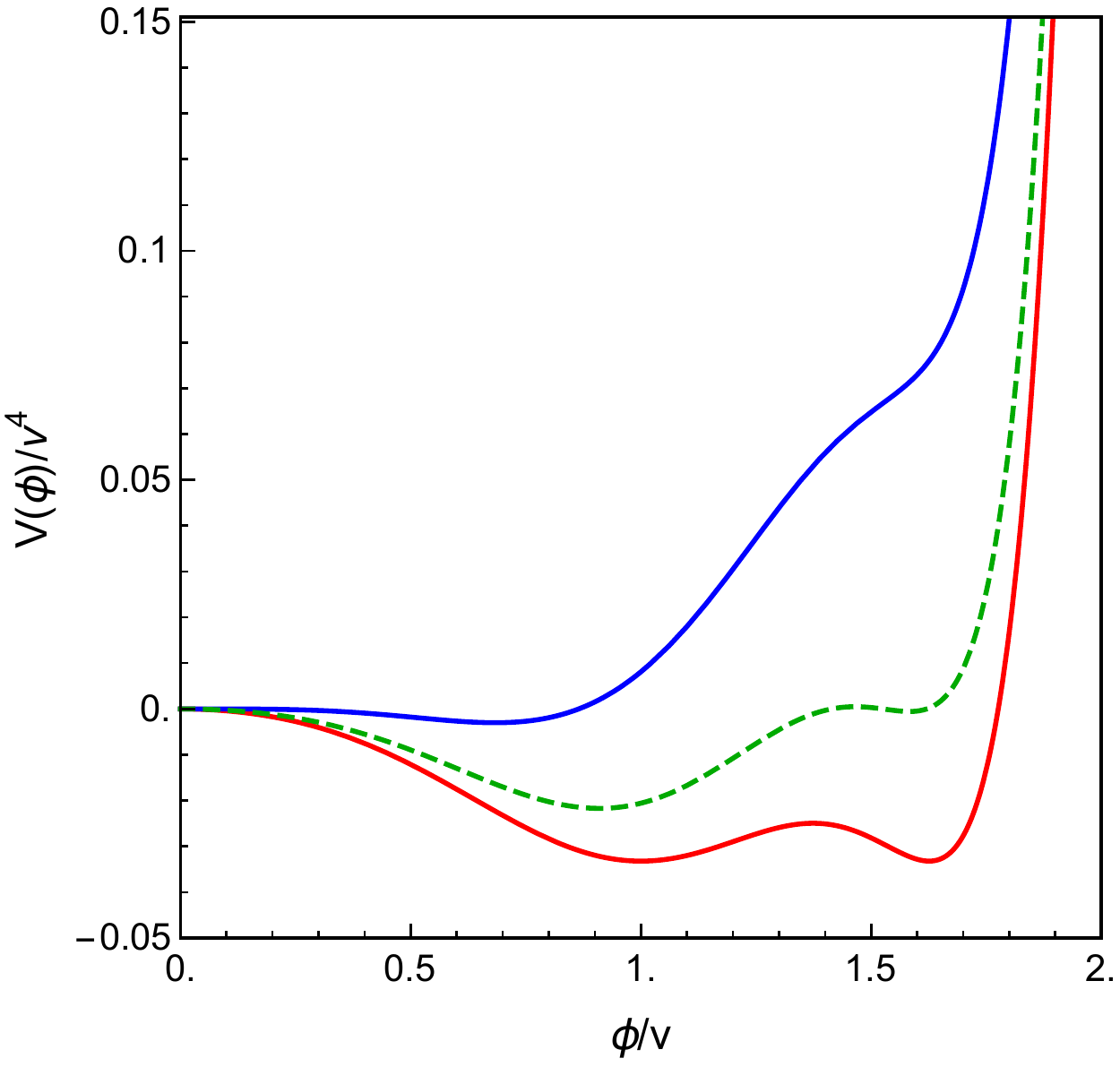}\quad
\includegraphics[width = 8.0cm, clip]{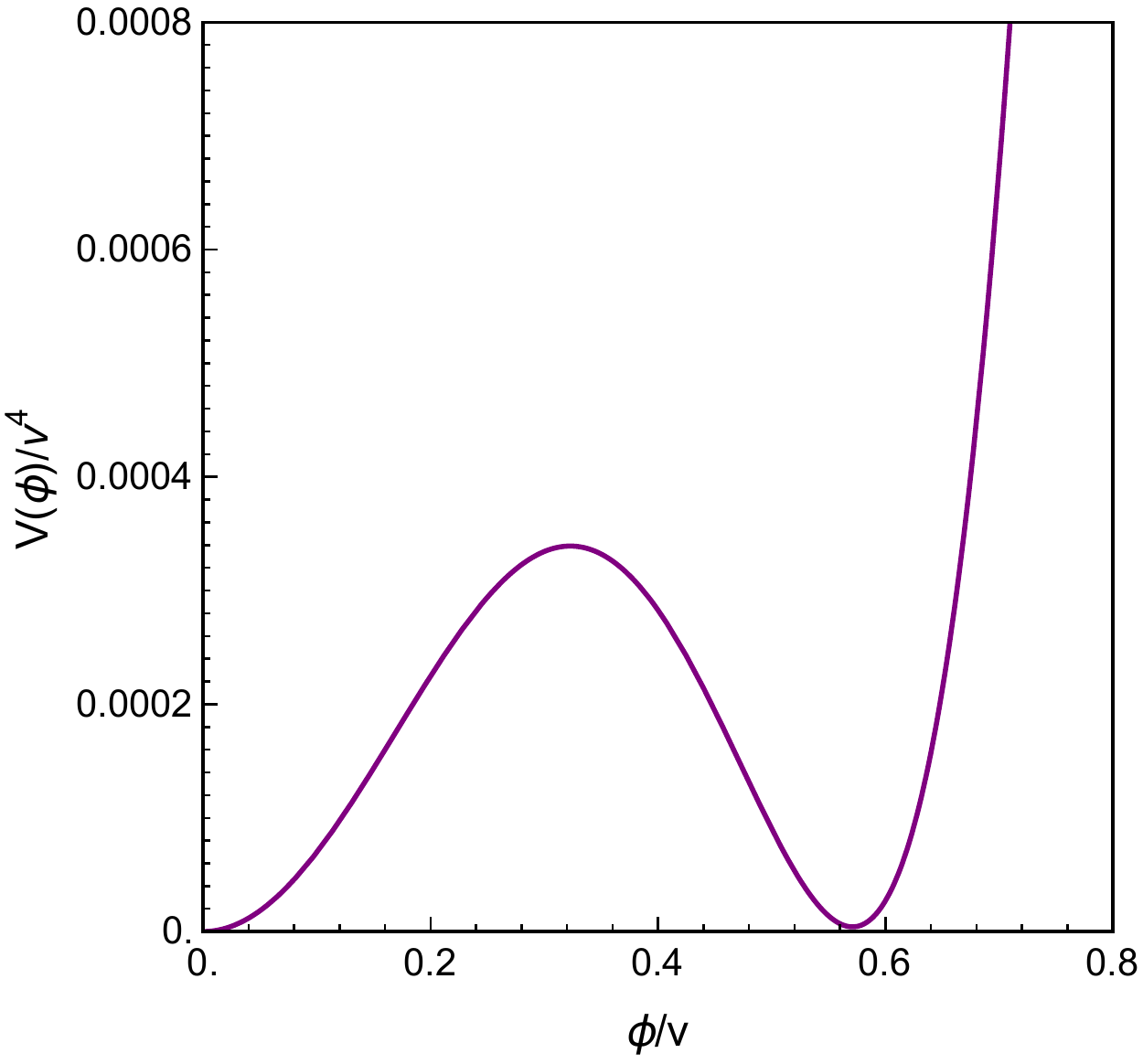}
\caption{Example of order $(\phi^\dagger\phi)^5$ potentials that correspond to the negative correction and also produce SFOEPT. In the left panel, the red line indicate the potential at $T=0$, the blue line correspond to the temperature where the curvature at $\phi=0$ is~0. The green line correspond to the intermediate temperature of~$\sim35$ GeV. The purple curve on the right shows the potential at $T=T_c$. 
The coefficients $c_6=0.906$, $c_8=-1$, $c_{10}=0.346$, while $\Lambda\sim263$~GeV, $T_c\sim44$ GeV assuming $a_0\sim3$ as in the SM and $\delta=-1.23$.}
\label{fig:phi10SFO}
\end{figure}

Fig.~\ref{fig:convergence}.c shows the possible modifications to $\lambda_3^{SM}$ by viable Higgs potentials that obey the experimental constraints on the Higgs mass and the VEV. We see that for the cut-offs near $250$ GeV, one can obtain variation in the $\lambda_3$ from $-5\lambda_5^{SM}$ to $7\lambda_3^{SM}$. Such large deviations make the triple Higgs coupling measurements at the LHC an exciting probe to the new physics. Fig.~\ref{fig:convergence}(d) shows a subset of the region in the left panel, in which a SFOEPT can take place. The black and the blue lines are retained from the Fig.~\ref{fig:convergence}.a and Fig.~\ref{fig:convergence}.b and serve as a reference for the comparison between the top and the bottom rows.

In Fig.~\ref{fig:convergence}.c and Fig~\ref{fig:convergence}.d the orange regions correspond to $|c_6|=1,\,|c_8|<1,\,0<c_{10}<1$, green region corresponds to $|c_6|<1,\,|c_8|=1,\,0<c_{10}<1$ and the purple region corresponds to $|c_6|<1,\,|c_8|<1,\,c_{10}=1$. As expected, two clusters are observed in the orange and green regions corresponding to the sign flips of $c_6$ and $c_8$ respectively. As in the case of $(\phi^\dagger\phi)^4$, there is overlap between the regions. The green region being present beneath all the area occupied by the orange region, while the purple region is present beneath all the area occupied by the other two colors.

An interesting feature of this kind of potential is the presence of negative enhancements in Fig.~\ref{fig:convergence}.d for the orange and green regions. This means that in principle
there are regions of parameters in which a negative enhancement of $\lambda_3$ may be obtained consistently with a FOEPT.  
Fig.~\ref{fig:phi10SFO} shows an example of the Higgs potentials, which is of  order $(\phi^\dagger\phi)^5$, and satisfies the Higgs mass and the VEV constraints and also undergo a SFOEPT with large negative enhancements of the triple Higgs coupling. In the left panel, the red line at the bottom corresponds to the potential at $T=0$, while the blue line depicts the potential at $T=T_f$ that corresponds to the curvature at $\phi=0$ being $0$. The green (dashed) line represents an intermediate temperature. In the right panel, the purple curve shows the phase transition of the corresponding potential in the left panel at $T=T_c$. Let us stress that negative enhancements of the triple Higgs couplings are only consistent with a FOEPT for small values of the cutoff, $\Lambda \simlt 350$~GeV. Hence, the correlation between the negative enhancements and the absence of a FOEPT remains generally valid.

\section{Minimal extension with a singlet}\label{sec:singlettotal}

Minimal extensions of the SM with just one singlet and their impact on electroweak baryogenesis have been studied in the literature~\cite{Profumo:2007wc,Barger:2007im,Wainwright:2012zn,Patel:2012pi,No:2013wsa,Profumo:2014opa,Gupta:2013zza,Barger:2011vm,
Noble:2007kk,Katz:2014bha}. Well motivated UV complete scenarios such as the NMSSM also have an additional singlet, which can mix with the SM Higgs~\cite{Menon:2004wv}. 

In subsection A we calculate the maximum enhancement of the triple Higgs coupling that can be allowed under the constraints of electroweak baryogenesis and the experimental constraints  coming from the LHC. In subsection B we assume that the singlet is heavy and integrate it out giving rise to an EFT. The resultant expressions for the triple Higgs enhancement and bounds on the FOEPT region can be shown to be the same as those generated from the full Lagrangian in the small mixing angle limit. At the same time, this approach represents an example of the potentials discussed in the previous section and therefore allows to discuss the validity and limitations of the effective theory approach.

\subsection{Enhancement in the full scalar Lagrangian of the singlet extension}\label{sec:fullRenorm}
Consider a general scalar potential, with one-loop thermal correction only in the mass term, that can be written in a canonically normalized Lagrangian for the SM extended with one singlet field $\phi_s$ 
\begin{align}
V(\phi_h,\phi_s,T)&=\frac{m_0^2+a_0T^2}{2}\phi_h^2+\frac{\lambda_h}{4}\phi_h^4+a_{hs}\phi_s\phi_h^2+\frac{\lambda_{hs}}{2}\phi_s^2\phi_h^2+t_s\phi_s+\frac{m_s^2}{2}\phi_s^2+\frac{a_s}{3}\phi_s^3+\frac{\lambda_s}{4}\phi_s^4\label{eqn:singletpot}
\end{align}
Here, $\phi_h$ is the higgs field. The VEV for the Higgs field is $v=246\,\,\text{GeV}$.   We assume that $m_s$ is larger than the weak scale and we therefore ignore the very small temperature corrections affecting the singlet mass.  

We stay in the limit, where $a_s$ and $\lambda_s$ are much smaller compared to $a_{hs}$ and $\lambda_{hs}$ and drop the $a_s$ and $\lambda_s$ terms. In this limit, we can retain analytical control over the expressions for the mixing and triple Higgs enhancement, which helps us clearly demonstrate the connection with the EFT. Within this approximation, the mass squared matrix in the basis $\left(\phi_h\,\,\phi_s\right)$ is
\begin{align}
\mathcal{M}^2&=\begin{pmatrix}m_{11}^2&m_{12}^2\\m_{21}^2&m_{22}^2\end{pmatrix}=\begin{pmatrix}2\lambda_hv^2&\quad2\left(a_{hs}+\lambda_{hs}v_s\right)v\\2\left(a_{hs}+\lambda_{hs}v_s\right)v&m_s^2+\lambda_{hs}v^2\end{pmatrix},\label{eqn:matrix}
\end{align}
where the VEV of the singlet field calculated at the Higgs vacuum is
\begin{align}
v_s=-\frac{t_s+a_{hs}v^2}{m_s^2+\lambda_{hs}v^2}.
\end{align}
The gauge eigenstate basis can be converted to the mass eigenstate basis as follows
\begin{align}
\phi_h&=\cos\theta\, h_1-\sin\theta\, h_2 + v\label{eqn:mixing00},\\
\phi_s&=\sin\theta\, h_1+\cos\theta\, h_2 + v_s \label{eqn:mixing0}.
\end{align}
The mixing is given as
\begin{align}
\tan\,2\theta=\frac{4v(a_{hs}+\lambda_{hs}v_s)}{2\lambda_hv^2-m_s^2-\lambda_{hs}v^2}&=\frac{4v(a_{hs}m_s^2-t_s\lambda_{hs})}{(2\lambda_hv^2-m_s^2-\lambda_{hs}v^2)(m_s^2+\lambda_{hs}v^2)}\label{eqn:mixing}
\end{align}

We use Equations~(\ref{eqn:matrix}) and~(\ref{eqn:mixing}), to convert the potential in Eq.~(\ref{eqn:singletpot}) to the mass basis $(h_2\,\,h_1)$ at the temperature $T=0$, where $h_1$ is the lighter of the two scalar fields. The third derivative of the potential in Eq.~(\ref{eqn:singletpot}) with respect to $h_1$ 
gives the triple Higgs coupling for the lower mass excitation as
\begin{align}
\lambda_3&=6\lambda_hv_h\cos^3\theta\left[1+\left(\frac{\lambda_{hs}v_s+a_{hs}}{\lambda_hv_h}\right)\tan\theta+\frac{\lambda_{hs}}{\lambda_h}\tan^2\theta\right].\label{eqn:singletlambda3I}
\end{align}
In the limit of $v^2 \ll m_s^2$, one can easily show that the $h_1$ mass is given by
\begin{equation}
m_{h}^2 = 2 \lambda_h v^2 -  4 v^2 \frac{(a_{hs} m_s^2 - t_s \lambda_{hs})^2}{(m_s^2 + \lambda_{hs} v^2)^3} 
\label{eqn:mh}
\end{equation}
Using Eq.~(\ref{eqn:singletlambda3}), Eq.~(\ref{eqn:mh}), and Eq.~(\ref{eqn:mixing}), we get
\begin{align}
\lambda_3&=\frac{3m_h^2}{v}\left[\cos^3\theta+\left(\frac{2\lambda_{hs}v^2}{m_h^2}\right)\sin^2\theta\cos\theta\right].\label{eqn:singletlambda3}
\end{align}
For $\theta=0$, we recover the SM result of $\lambda_3=\frac{3m_h^2}{v}$. 

In the small $\theta$ limit, the above formula reduces to 
\begin{align}
\lambda_3&=\frac{3m_h^2}{v}\left[1+\left(\frac{2\lambda_{hs}v^2}{m_h^2}-\frac{3}{2}\right)\tan^2\theta\right].\label{eqn:singletlambda3smalltheta}
\end{align}

The same result can be recovered in the EFT approach by integrating out the heavier state as shown in the next section~\ref{sec:singletEFT}. For the FOEPT in such a potential, we impose the following conditions.
\begin{align}
V\left(0,T_c\right)&=V\left(v_c,T_c\right),\quad V'\left(v_c,T_c\right)=0.
\end{align}
This leads to~\cite{Menon:2004wv}
\begin{align}
v_c^2 = \frac{1}{\lambda_{hs}}\left(-m_s^2 + \sqrt{\frac{2}{\lambda_h}}\,\left|m_s\,a_{hs}-\frac{\lambda_{hs}\,t_s}{m_s}\right|\right).\label{eqn:phic}
\end{align}
Here $v_c$ is the value of the doublet scalar field at the critical Temperature ($T_c$). The value of $S$ is set to 
\begin{align}
v_{s,c}&=-\frac{t_s+a_{hs}v_c^2}{m_s^2+\lambda_{hs}v_c^2},
\end{align}
which minimizes the potential at $\phi_h=v_c$. The constraints on the derivatives
\begin{align}
V'\left(\phi_c,T_c\right)&=0,\quad V'\left(v,0\right)=0 ,\label{eqn:tc}
\end{align}
imply $a_0T_c^2 = 8\left(F(v_c^2) - F(v^2)\right)$. Here $F(\phi^2)=-\frac{V'(\phi,0)}{\phi}\text{ and }v=246\text{ GeV}$. 

\begin{figure}[singlet]{
\includegraphics[width = 7.2cm, height=7.2cm, clip]{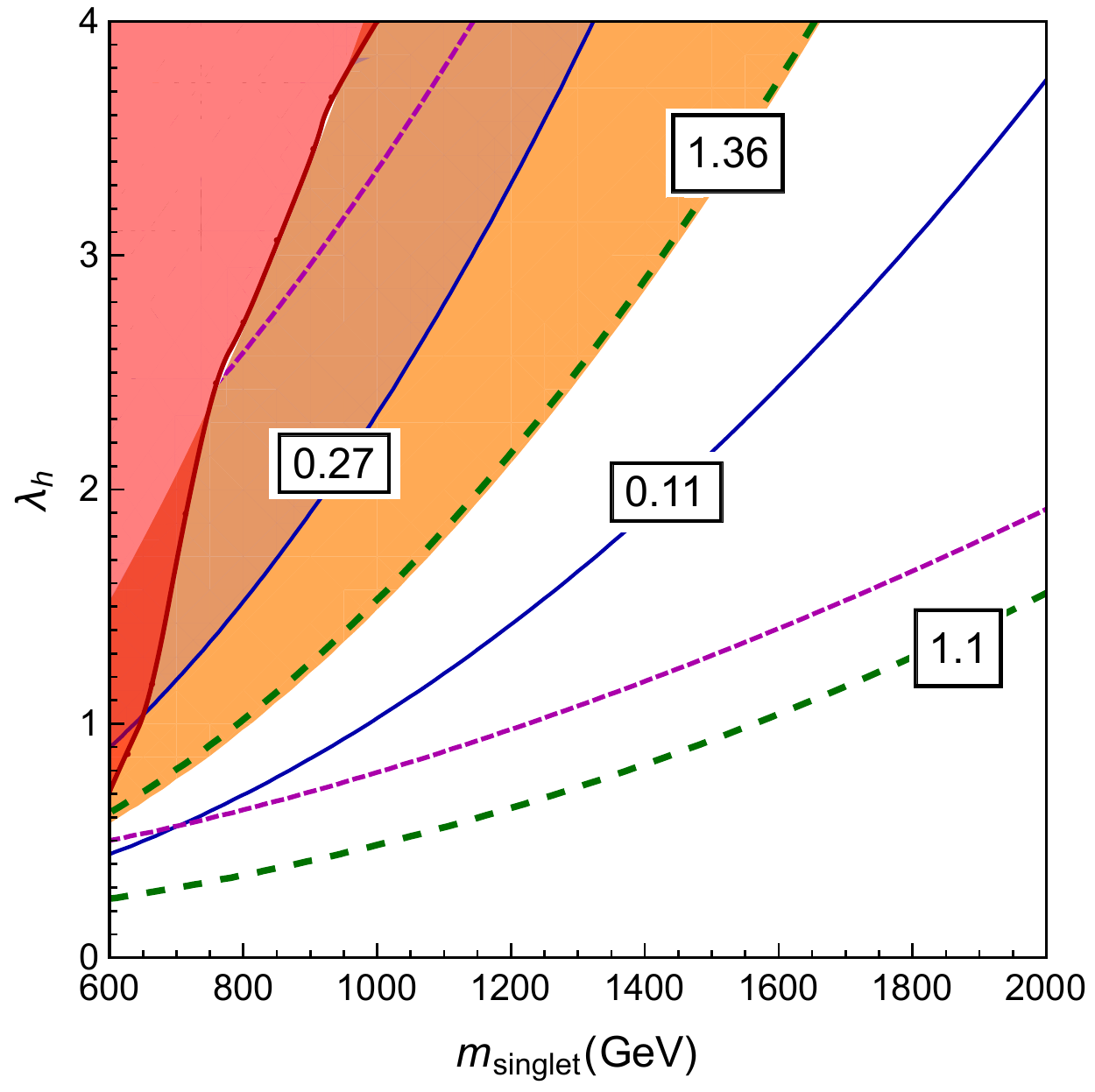}
\includegraphics[width = 7.2cm, height=7.2cm, clip]{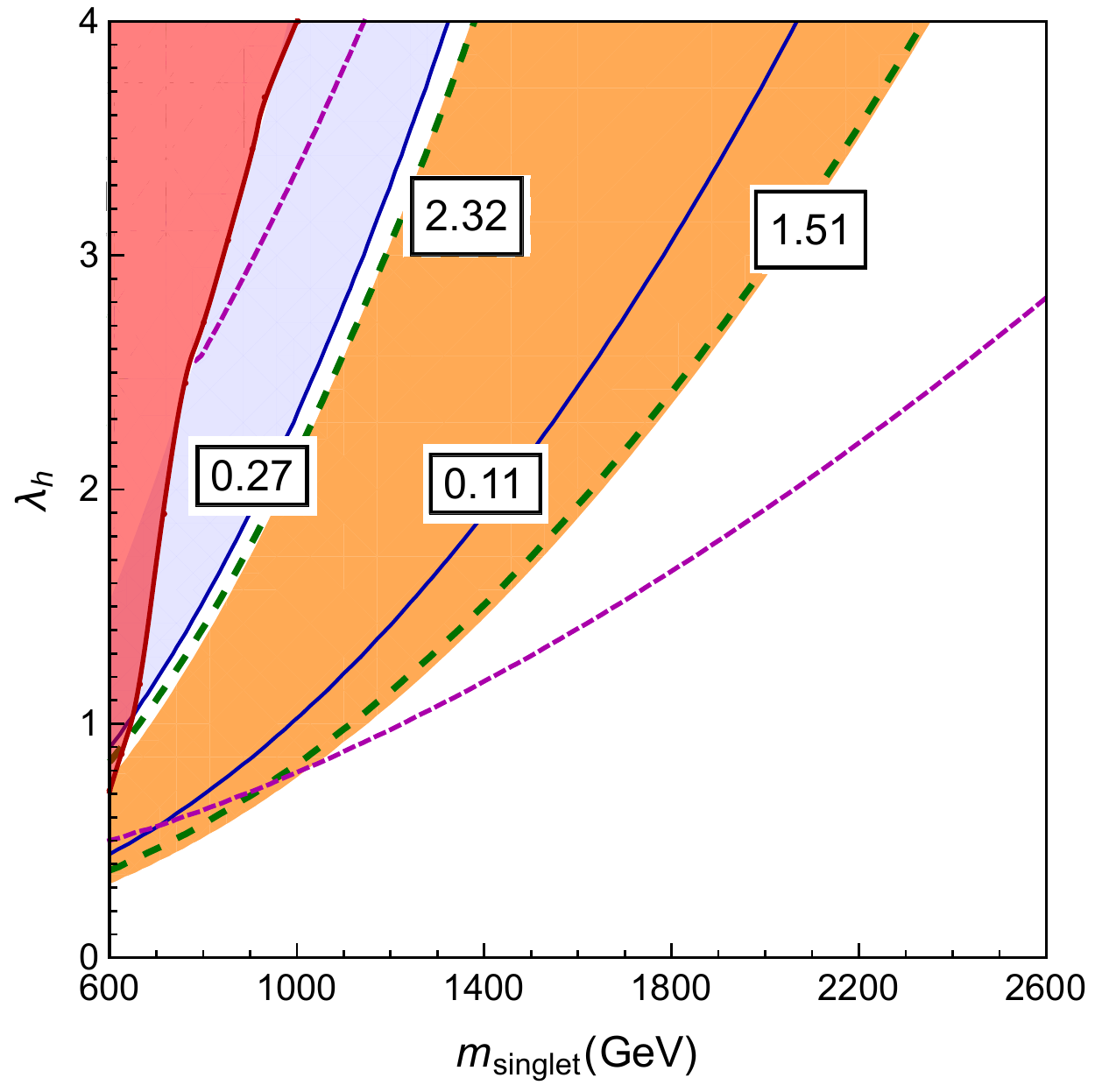}
\includegraphics[width = 7.2cm, height=7.2cm, clip]{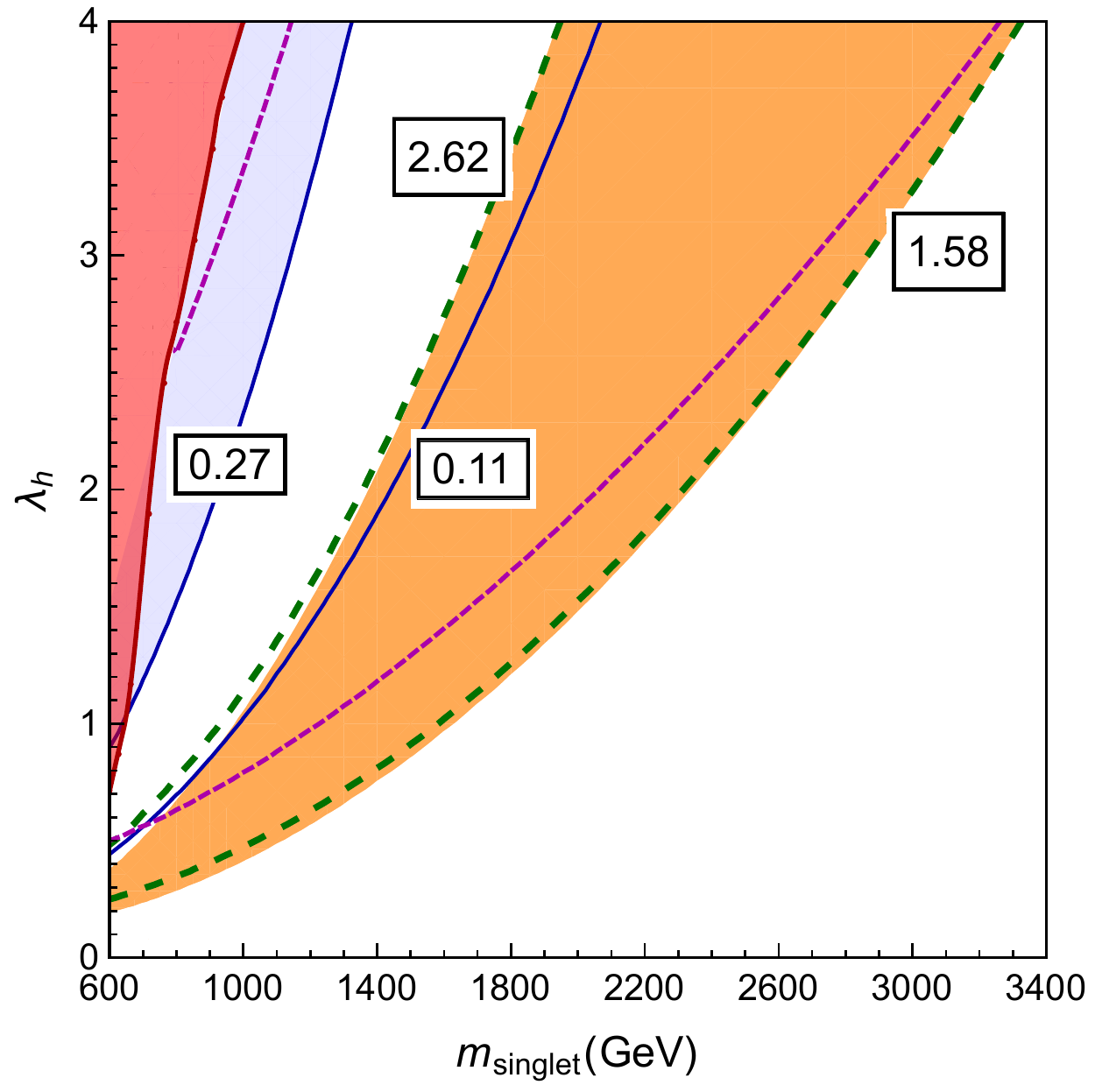}
\includegraphics[width = 7.2cm, height=7.2cm, clip]{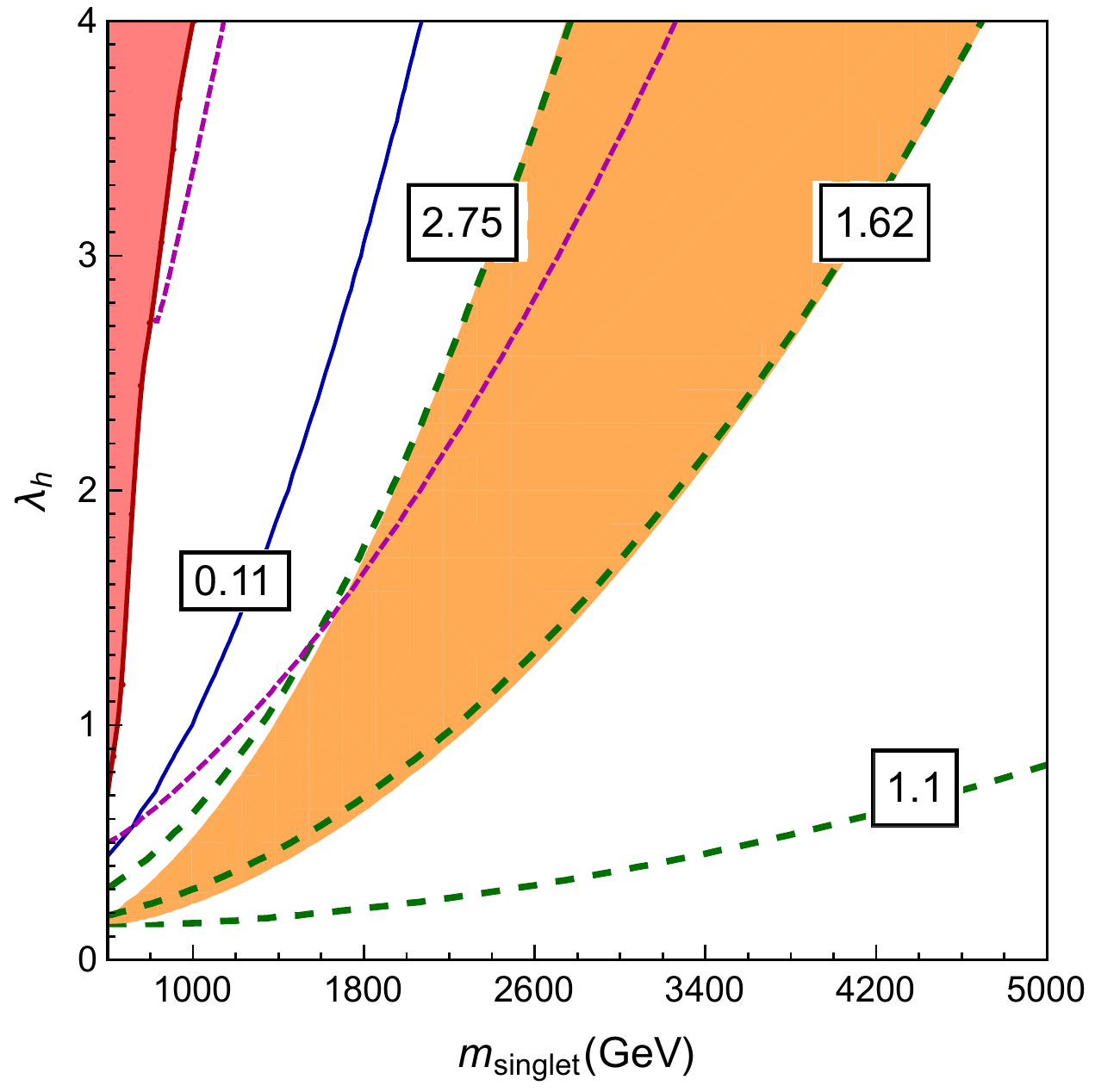}
\caption{
Contours of the mixing parameter $\sin^2\theta$ (solid blue line) and of the enhancement of the triple-Higgs coupling (dashed green line) given by Eq.~(\ref{eqn:singletlambda3}) in the $m_{\rm singlet}$--$\lambda_h$ plane. Blue shaded region denotes $2\sigma$ exclusion due to gluon fusion channel. The orange shaded region represents the region consistent with a FOEPT. The region excluded up to $2\sigma$ confidence level by Higgs precision measurements is shaded in red. The constraints coming from $m_W$ are shown by magenta (short-dashed) lines. In the top-left panel we present results for  $\lambda_{hs}=0.5$, while in the top-right, bottom-left and bottom-right panels we present results for $\lambda_{hs}=1,\,2,\,4$ respectively.}
\label{fig:msing}}
\end{figure}

In Fig.~\ref{fig:msing} we show the enhancements of the trilinear couplings for different
values of the singlet mass $m_{\rm singlet}$ and the quartic coupling $\lambda_h$. 
 The orange region in the Fig.~\ref{fig:msing} corresponds to the region consistent with a FOEPT, i.e. the boundaries correspond to $v_c^2=0$ and $T_c^2=0$. 

From Eq.~(\ref{eqn:phic}) and Eq.~(\ref{eqn:mh}), it follows that for $T_c = 0$,
or equivalently $v_c = v$ one obtains
\begin{equation}
\tan^2\theta (v_c = v) \simeq \frac{m_h^2}{\lambda_{hs} v^2}
\end{equation}
Similarly, for $v_c = 0$, one obtains
\begin{equation}
\tan^2\theta (v_c = 0) \simeq \frac{m_h^2}{3 \lambda_{hs v^2}}
\end{equation}

Using these expressions for small mixing angles,  Eq.~(\ref{eqn:singletlambda3smalltheta}), 
one can easily show that
\begin{equation}
\delta (v_c = v) \simeq  2 - \frac{3 \ m_h^2}{2 \lambda_{hs} v^2}
\end{equation}
while in the case of $v_c = 0$ one obtains
\begin{equation}
\delta (v_c = 0) \simeq \frac{2}{3} - \frac{m_h^2}{ 2 \lambda_{hs} v^2}.
\end{equation} 

The region compatible with a FOEPT is always between these boundaries of $v_c=0$ and $v_c=v$. Thus, the enhancement to the triple Higgs coupling is always less than 3, a result
similar to the one obtained in the $(\phi^{\dagger}\phi)^3$ extension of the Higgs potential
discussed in section~\ref{sec:phi6}.  
Finally, let us mention that the SFOEPT constraint of $v_c>0.6\,T_c$, is almost always satisfied in the showed orange region.

In Fig.~\ref{fig:msing}, we also show experimental constraints coming from Higgs physics and electroweak precision measurements. The mixing parameter $\sin^2\theta$ is denoted by the blue contours. The precision measurements of the SM-like Higgs properties at the LHC already impose strong constraints on the possible mixing angle of the singlet with the doublet. For example, the measurement of the Higgs production signal rates imposes an upper bound  on $\sin^2\theta$.  If one takes the gluon fusion production process, the
combined measurement of ATLAS and CMS gives a signal strength~\cite{ATLASCMS}
\begin{equation}
\mu_{\rm ggF} = 1.03^{+ 0.17}_{ - 0.15}.
\label{eq:muggF}
\end{equation}
When the other subleading processes, including the weak boson fusion, associated production and $tth$ production are considered z, one obtains a combined signal strength
\begin{equation}
\mu  = 1.09^{+ 0.11}_{ - 0.10}. 
\label{eq:mutot}
\end{equation}
Since the mixing with a singlet leads to an overall decrease of all couplings to fermions and gauge bosons, the Higgs decay branching ratios will not be affected and the signal strength will be proportional to $\cos^2\theta$. Hence, from Eqs.~(\ref{eq:muggF}) and (\ref{eq:mutot}) one obtains a $95\%$ confidence level upper bound on $\sin^2\theta$, namely
\begin{equation}
\sin^2\theta < 0. 11
\end{equation}
if the fit to all production processes is considered, and $\sin^2\theta < 0.27$ if only the more precisely measured gluon fusion processes are considered.  In our work, we shall considered both bounds, as an indication of the constraints on the possible realization of this scenario.  

In the case of small $\theta$, as seen from the Eq.~(\ref{eqn:singletlambda3smalltheta}),
the correction to $\lambda_3$ compared to the SM is proportional to $\sin^2\theta$. From this, it is evident that the upper bound on the mixing will be translated into an upper bound on the enhancement
of $\lambda_3$, 
\begin{equation}
\delta < \sin^2 \theta_{\rm max} \left( \frac{2 \lambda_{hs} v^2}{m_h^2} - \frac{3}{2} \right) 
\sim \sin^2 \theta_{\rm max}  \left( 8 \lambda_{hs} - \frac{3}{2} \right).
\label{eqn:maxmix}
\end{equation}
Hence, these
constraints become more severe for smaller values of $\lambda_{hs}$. 

From Eqs.~(\ref{eqn:singletlambda3smalltheta}) and (\ref{eqn:maxmix}), we also see that reducing $\lambda_{hs}$ below $\frac{3 m_h^2}{4v^2}$ leads to small negative  values of $\delta$. Therefore, a small suppression of the triple higgs coupling with respect to the SM is viable for these values of $\lambda_{hs}$. As shown in  Fig.~\ref{fig:msing}, for these values of $\lambda_{hs}$ the FOEPT region shifts rapidly to the higher mixing values and becomes unviable. Thus, there is trade-off between FOEPT and supression of the triple Higgs coupling with respect to the SM as shown in the EFT case in the previous section.

Moreover, a light singlet that mixes with the SM Higgs will be produced at the LHC and may be searched for in various decay channels.
This puts an additional constraint on the realization of this model, which is also shown in Fig.~\ref{fig:msing}. The region to the left of the dark red solid line is excluded by the Higgs searches in the WW and ZZ channels~\cite{Khachatryan:2015cwa}. 

The mixing between the doublet and the singlet is also constrained by precision W mass measurement~\cite{Robens:2015gla,Lopez-Val:2014jva}. The world average for the mass of the W boson is $m_W~ =~ 80.385~ \pm ~0.015$~GeV~\cite{Wmass} including data from LEP II~\cite{LEP}, CDF~\cite{CDF} and D0~\cite{D0}. The prediction of the W mass is obtained by calculating the muon life time, which yields the relation, 
\begin{equation}
m_W^2(1-\frac{m_W^2}{m_Z^2}) = \frac{\pi \alpha}{\sqrt{2} G_F}(1+\Delta r),
\label{eq:mW}
\end{equation}
where $\Delta r$ summarizes the radiative corrections. In the SM, $m_W^{SM} = 80.361 \pm 0.007$ GeV~\cite{PhysRevD.22.2695, Awramik:2003rn}, which corresponds to  $\Delta r^{SM} = (37.979 \pm 0.406)\times 10^{-3}$ , with the mass of the Higgs $m_h = 125$ GeV. From Eq~(\ref{eq:mW}), $\Delta r^{exp} = (36.32 \pm 0.96) \times 10^{-3}$, which is about 1.7$\sigma$ from the SM value.  $\Delta r$ can be parametrized as 
\begin{equation}
\Delta r = \Delta \alpha + \frac{c_w^2}{s_w^2}(\frac{\delta m_Z^2}{m_Z^2} - \frac{\delta m_W^2}{m_W^2}) + (\Delta r)_{rem} ,
\end{equation}
where $\Delta \alpha$ is the radiative correction to the fine structure constant $\alpha$, and $c_w$ and $s_w$ are the cosine and sine of the weak mixing angle. The second term is the on-shell self-energy correction to the gauge boson masses, which is well approximated by its value at zero momenta, and relates to the $\rho$ parameter as  $ -\frac{c_w^2}{s_w^2} \Delta \rho$. The last term, $(\Delta r)_{rem}$, includes vertex corrections and box diagrams at one loop level, which are subleading. In the case of having a singlet mixed with the SM Higgs, $\Delta r$ is
given by
\begin{equation}
\Delta r = \Delta r^{SM}  -\frac{c_w^2}{s_w^2} (\Delta \rho^{\rm singlet} - \Delta \rho^{SM}),
\label{eq:r}
\end{equation} 
where $\Delta \rho^{\rm singlet}$ and $\Delta \rho ^{SM}$ are the $\Delta \rho$ calculated in the the case with a mixed-in singlet and the SM~\cite{Hagiwara:1994pw}. 
\begin{equation}
\Delta \rho ^{\rm singlet} - \Delta \rho ^{SM} = G_F\frac{m_Z^2}{2\sqrt{2}\pi^2} sin^2\theta \left(H_T (\frac{m^2_{Singlet}}{m_Z^2} ) - H_T (\frac{m^2_{h}}{m_Z^2} )\right), 
\end{equation}
where
\begin{equation}
 H_T(x) = \frac{3}{4} x \left(\frac{\log(x)}{1-x} - \frac{\log(x\times m_Z^2/m_W^2)}{1-x \times m_Z^2/m_W^2} \right).
\end{equation}
The constraints on $\sin^2\theta$ obtained from the $W$ mass become quite severe since
as mentioned above, the SM is already in tension with the $W$ mass measurement, and the
singlet contribution increases this tension.  The 2~$\sigma$ constraint coming from $\Delta r$
calculated from Eq~(\ref{eq:r}) is shown by the lowered dashed magenta
line in Fig.~\ref{fig:msing}.   On the other hand, if one assumes that some other new physics, which does not modify the loop induced Higgs production processes in a relevant way is responsible for the difference between the SM and the current W mass measurement the  bounds become significantly weaker as seen from the upper dashed magenta line in Fig.~\ref{fig:msing}. It follows from Fig.~\ref{fig:msing} that even considering the tight constraints coming from Higgs measurements and precision electroweak parameters, a strongly first order phase transition is possible in these scenarios, provided $\lambda_{hs} \simgt 1$. Large values of the singlet mass, of the order of the TeV scale, are possible in this case, making $\sin^2\theta$ small. In our analysis, we ignore the one loop contributions to the effective potential since they are suppressed compared to the tree level mixing effects. When $\lambda_{hs}$ is sizable, as we show in the lower panels in Fig.~\ref{fig:msing}, those corrections may not be negligible and should be taken into account in a more refine analysis of the critical parameters.

Before concentrating on the EFT analysis let us stress that an important contribution to the double
Higgs production cross section that is always missed in this analysis is the 
resonant double Higgs production induced by the singlet. This can lead to a relevant contribution
if the singlet is below the TeV scale and the mixing is sizable~\cite{Chen:2014ask}. For instance, at the LHC with
a center of mass energy of 14~TeV a 500 GeV singlet with a mixing of 
$\sin^2\theta = 0.2$, will lead to a resonant production cross section through gluon fusion for the singlet of about 1.13 pb~\cite{LHCHXSWG}. Under these conditions the branching ratio BR($S \rightarrow hh$) $\sim$ 0.013. Then the double Higgs production rate induced by the singlet is about 15~fb, which is about a factor of 4 smaller than the SM double Higgs production rate. Such a singlet would show up in the invariant mass distribution as a narrow resonance, as the singlet width is about 17~ GeV. When the singlet gets heavier, say about 1~TeV, and for a mixing angle
$\sin^2\theta = 0.1$, the double Higgs production induced by the singlet is reduced to about 2.6~fb, which is significantly suppressed compared to the double Higgs production from the box and triangle diagrams, and difficult to detect in the standard decay channels.  Then, in the region of a heavy singlet and small mixing angle,  the EFT gives a proper description of the physics involved in double Higgs production. In
this case, the singlet presence may only be inferred indirectly and one can make contact
with an effective theory description of the modification of the trilinear couplings and of the
double Higgs production rate.

\subsection{EFT formulation for the singlet extension}\label{sec:singletEFT}
In the limit of large values of the singlet mass $m_s$,  and
small mixing between the SM-like Higgs and the heavy singlet, we can integrate out the heavy singlet, and the resulting EFT should describe the same physics as we have described in
the previous subsection. 

For momenta very small compared to the masses of the scalars, solving the equation of motion for the singlet gives
\begin{align}
\phi_s=-\frac{t_s+a_{hs}h^2}{m_s^2+\lambda_{hs}h^2}.\label{eqn:EOM}
\end{align}
Substituting this into the original potential in Eq.~(\ref{eqn:singletpot}) yields an effective
potential for $h$, which is given by~\cite{Menon:2004wv}
\begin{align}
V(h,T)&=\frac{m^2(T)}{2}\phi_h^2+\frac{\lambda_h}{4}\phi_h^4-\frac{\left(t_s+a_{hs} \phi_h^2\right)^2}{2\left(m_s^2+\lambda_{hs}\phi_h^2\right)}.\label{eqn:effsingtpot}
\end{align}
where $m^2(T) = m_0^2+a_0T^2$.
The integration out of the singlet also leads to a modification of the Higgs kinetic
term, which means that the well normalized Higgs field $H$ will no longer be given by $h$,
but will be affected by the mixing with the singlet. In other words, 
substituting the EOM of $S$ in its kinetic term leads to an $h$ dependent normalization factor,
\begin{align}
(\partial_\mu \phi_h)(\partial^\mu \phi_h)+(\partial_\mu \phi_s)(\partial^\mu \phi_s)\rightarrow\left(1+{\frac{4\,\phi_h^2(am_s^2-t_s\lambda_{hs})^2}{(ms^2 + \lambda_{hs} \phi_h^2)^4}}\right)(\partial_\mu \phi_h)(\partial^\mu \phi_h).
\end{align}

Demanding $H$ to be well normalized  and retaining up to first order in the small parameter 
\begin{equation}
z =\frac{(am_s^2-t\lambda_{hs})^2v^2}{m_s^8}
\end{equation}
we obtain
\begin{align}
\phi_H=\phi_h+\frac{2 z \phi_h^3}{3\,v^2}+\mathcal{O}(\phi_h^5).
\label{eq:firstHhrel}
\end{align}

The corresponding $c_H$ is 
\begin{equation}
\frac{c_H}{4\Lambda^2} = \frac{z}{v^2}.
\label{eqn:cH}
\end{equation}
The variable $z$ defined above is related to the mixing angle between the singlet and the
doublet. From Eq.~(\ref{eqn:mixing}), we can write
\begin{align}
\tan^22\theta&=\frac{16z}{\left(2\lambda_hy-1-\lambda_{hs}y\right)^2\left(1+\lambda_{hs}y\right)^2}=4\tan^2\theta+\mathcal{O}(\tan^3\theta).
\end{align}
Substituting Eq.~(\ref{eqn:lambdah}) and retaining first order in z we get
\begin{align}
\tan^22\theta&=16z+\mathcal{O}(z^2)=4\tan^2\theta+\mathcal{O}(\tan^3\theta)
\implies \tan^2\theta\sim4z
\end{align}

Inverting the relation between $\phi_h$ and $\phi_H$ given in Eq.~(\ref{eq:firstHhrel}) one obtains
\begin{align}
\phi_h= \phi_H-\frac{2z}{3\,v^2}\phi_H^3+\mathcal{O}(\phi_H^5),
\label{eq:phihphiH}
\end{align}
Substituting this in Eq.~(\ref{eqn:effsingtpot}), we get an effective potential, which retaining up to order $H^6$ corrections is given by
\begin{align}
V_{eff}(\phi_H,T)&=\frac{m^2}{2}\phi_H^2 +\left(\frac{\lambda_h - 2 z/y}{4}-\frac{2 m^2 z}{3 v^2}\right) \phi_H^4 + \left(\frac{- 4 z (\lambda_h- 2z/y) + 3 z \lambda_{hs}}{6 v^2}\right) \phi_H^6,\label{eqn:singleteffpot}
\end{align}
where $y = v^2/m_s^2$. This shows that the presence of a large negative correction to the quartic coupling, of order $2 z/y$. This correction, which depends only on ratios of mass parameters, allows for the presence of a negative effective quartic coupling which according to our analysis of the EFT at this order in section~\ref{sec:phi6}, is essential for the obtention of a FOEPT. 

Using this potential Eq.~(\ref{eqn:effsingtpot}) we apply the Higgs mass condition to write
\begin{align}
\left(V_{eff}''-\frac{V_{eff}'}{\phi_H}\right)\bigg|_{\phi_H=\left<\phi_H\right>}=m_H^2,\quad\text{where}\quad \left<\phi_H\right>=v+\frac{2zv}{3}.
\label{eqn:omega}
\end{align}
Solving this simultaneously with 
\begin{align}
\frac{V_{eff}'}{\phi_H}\bigg|_{\phi_H=\left<\phi_H\right>}=0,
\end{align}
leads to a relation of the value of $\lambda$ and the Higgs mass. 
\begin{align}
\lambda = \lambda_h - \frac{2 z}{y} =\frac{m_H^2}{2 v^2} + \left(\frac{2\,m_H^2}{v^2}-6 \lambda_{hs}\right)z.\label{eqn:lambdah}
\end{align}
Since $m_H^2/(2 v^2) \simeq 1/8$, for small values of $z$ the coefficient of the quartic coupling $\lambda$ is small in magnitude and may be negative for $\lambda_{hs}$ of order 1.

Moreover, a sizable
correction to the sixth order term appears, 
which is there even in the absence of kinetic terms corrections. Observe that $\lambda_h - 2z/y$, which as shown above corresponds to $\lambda$ in the EFT analysis, appears also in the first term in the $\phi_H^6$ coefficient. Since $\lambda$ is small as discussed above, the $\phi_H^6$ coefficient is dominated by the second term in the bracket.  The cut off scale can be then calculated from 
\begin{equation}
\frac{c_6}{8\Lambda^2} \sim \frac{3\lambda_{hs}z}{6v^2} = \frac{\lambda_{hs}(a m_s^2-t \lambda_{hs})^2}{2 m_s^8}.
\label{eqn:c6}
\end{equation}
The corresponding cutoff scale is, for $c_6 = 1$
\begin{equation}
\Lambda^2 = \frac{m_s^8}{4\lambda_{hs}(a m_s^2 - t\lambda_{hs})^2}.
\end{equation}
Thus, when $(a m_s^2 - t\lambda_{hs})$ and $\lambda_{hs}$ become sizable, $\Lambda$ could be significantly lower than $m_s$. However, $a m_s^2 - t\lambda_{hs}$ is related to $\sin^2\theta$, which is constrained by electroweak symmetry breaking, precision Higgs measurements, heavy SM-like Higgs searches, and W mass as discussed above, the cutoff scale can not be lowered arbitrarily. For example, since $\lambda$ is small, from Eq.~(\ref{eqn:lambdah}), we have
\begin{equation}
\lambda_h \sim \frac{2z}{y} = \frac{2(a m_s^2 - t \lambda_{hs})^2}{m_s^6}.
\end{equation}
Then the cutoff scale is about
\begin{equation}
\Lambda^2 \sim \frac{m_s^2}{2\lambda_h\lambda_{hs}.}
\label{eqn:ctff}
\end{equation}
It is instructive to compare these results with those shown in Fig.~\ref{fig:msing}.
For instance, when $m_{\rm singlet}$ is about 1.4 TeV, $\lambda_{hs} = 2$, and $\lambda_h = 2$,  that is close to the boundary of the orange region in 
the bottom-left panel of Fig.~\ref{fig:msing}, the cutoff scale is about 494~GeV, which is about the lower bound of the cutoff scale in a $(\phi^\dagger\phi)^3$ theory and is consistent with the left boundary of the orange region in this figure.  Similarly,
for the same results of $\lambda_h$ and $\lambda_{hs}$, and for $m_{\rm singlet} = 2.4$~TeV, that is closed to the other boundary in the bottom-left panel of Fig.~\ref{fig:msing}, the effective cutoff scale that is obtained from Eq.~(\ref{eqn:ctff}) is about 848~GeV that is very close to the upper bound on $\Lambda$ that is obtained for a FOEPT in the $(\phi^{\dagger} \phi)^3$ extension.  One can
check that similar values of the cutoff are obtained at the left and right boundaries of the orange
regions in Fig.~\ref{fig:msing} for other values of $\lambda_h$, $\lambda_{hs}$ and $m_{\rm singlet}$. 

After
substituting Eq.~(\ref{eqn:lambdah}) and considering the field fluctuations of the field $\phi_H$,
\begin{equation}
\phi_H = v_H + H,
\end{equation}
we obtain,
\begin{align}
\lambda_3 \equiv g_{HHH} &=\frac{3m_H^2}{v}\left(1+4\,z\,\left(\frac{2\lambda_{hs}v^2}{m_H^2} - \frac{3}{2}\right)\right).\label{eqn:enhwithz}
\end{align}

Using this in Eq.~(\ref{eqn:enhwithz}) we obtain
\begin{align}
\lambda_3&=\frac{3m_H^2}{v}\left(1+\left(\frac{2\lambda_{hs}v^2}{m_h^2} - \frac{3}{2}\right)\tan^2\theta\right)
\label{eqn:enhncmnt}
\end{align}

This formula is the same as that obtained in Eq.~(\ref{eqn:singletlambda3smalltheta}) from the small mixing limit of the enhancement up to $\tan^2\theta$ order in the full renormalizable Lagrangian. Thus, as expected, the EFT approach is equivalent to the small mixing limit of the full theory.   To make the analogy more transparent let's emphasize that from Eq.~(\ref{eq:phihphiH}) 
the fluctuations of the
field $\phi_h = v + h$ and $H$  are related by
\begin{equation}
h = \left(1 - \frac{\tan^2\theta}{2} \right) \ H \simeq \ \cos\theta \ H
\end{equation}
That is the same relation we obtain between $h_1$ and $h$ in the full theory, 
Eq.~\ref{eqn:mixing00}, when we consider negligible $h_2$  fluctuations associated with its decoupling from the low energy theory. 

We note that the effective potential derived in Eq.~(\ref{eqn:singleteffpot}) is of order $\phi_H^6$. This is the same order as the $(\phi^\dagger\phi)^3$ potential described in section~\ref{sec:phi6}. In this case, however, the range of values of $\delta$ is not constrained from $2/3$ to $2$ as expected from the $(\phi^\dagger\phi)^3$ theory,  but is shifted to lower values. This is due to the kinetic terms corrections we were not considered in the analysis in Section~\ref{sec:EWPT}.  For $\lambda_{hs} \simgt 1$,
the kinetic term corrections remain significantly smaller than the ones associated with the effective potential modification, which are controlled
by the $\lambda_{hs}$ coupling.
 Expressing Eq.~(\ref{eqn:enhwithz}) in terms of $c_6$ and $c_H$, using Eq.~(\ref{eqn:c6}) and Eq.~(\ref{eqn:cH}), we obtain
 \begin{equation}
 \lambda_3 = \frac{3m_H^2}{v} \left(1+c_6 \frac{2 v^4}{m_h^2 \Lambda^2}-\frac{3}{2}c_H\frac{v^2}{\Lambda^2} \right),
 \end{equation}
 This is consistent with Eq.~(\ref{eqn:enh6}) when $c_H = 0$. Also, this is consistent with Eq.(34) in Ref~\cite{Gupta:2013zza} and Eq.~(124) of Ref.~\cite{Giudice:2007fh} when taking $\lambda = m_h^2/(2 v^2)$.  As mentioned before, our expression is more  suitable for the study of the region of parameters consistent with a FOEPT in which $\lambda$ is small and negative and the proper relation between $\lambda$ and the Higgs mass can only be obtained after including the higher order corrections proportional to $c_6$, Eq.~(\ref{eqn:curvature}).  

  
Higher powers of $\phi_H$ in the Eq.~(\ref{eqn:singleteffpot}) can be obtained by retaining more terms in the expansions with respect to $z$ and $y$ variables.  For instance, we
have checked that at next order the well normalized field is given by
\begin{align}
\phi_H &= \phi_h+\frac{2 z \phi_h^3}{3 v^2}-\frac{2(z^2 + 4 y z \lambda_{hs}) \phi_h^5}{5 v^4}
\end{align}
Expressing $h$ in terms of $H$,
\begin{align}
\phi_h &= \phi_H-\frac{2 z \phi_H^3}{3 v^2}+\frac{2 (13 z^2 + 12 y z \lambda_{hs}) \phi_H^5}{15 v^4}
\end{align}
one can obtain the value of $\delta$, as we did below, that is given by
\begin{align}
\delta &= \left(-6+\frac{8v^2\lambda_{hs}}{m_h^2}\right)z + \left(30-\frac{48v^2\lambda_{hs}}{m_h^2}\right) z^2 + \left(40\lambda_{hs}-\frac{32v^2\lambda_{hs}^2}{m_h^2}\right)yz 
\end{align}
That indeed reproduces the small $\theta$ expansion of the exact formula, Eq.~(\ref{eqn:singletlambda3smalltheta}).  

Again, it is straightforward to see that $H$ and $h$ are related by
\begin{equation}
h =  \left( 1 - \frac{\tan^2\theta}{2} + \frac{3 \tan^4\theta}{8} \right) \ H \simeq \
\cos\theta \ H
\end{equation} 
as expected from the relation between $h_1$ and $h$ in the full theory, Eq.~(\ref{eqn:mixing00}).


Before we concentrate on collider phenomenology, let us comment on the negative enhancement in a theory with a mixed-in singlet. Once a small singlet quartic coupling $\lambda_{s}$ is turned on to stabilize the potential, $\lambda_{hs}$ can go to negative values, as long as $|\lambda_{hs}| < \sqrt{\lambda_h \lambda_{s}}$. A small $\lambda_{s}$ leads to a contribution to $\lambda_3$ suppressed by $\sin^3 \theta$, $\sim 6 \lambda_{s} v_s \sin^3 \theta$.  As seen in Eq.~(\ref{eqn:enhncmnt}), a negative $\lambda_{hs}$ provokes a negative enhancement while a small positive $\lambda_s$ adds negligible contribution to $\lambda_3$. We note that, in the EFT context,  the $\lambda_s$ term generates a term of order $\frac{1}{4} \lambda_s \frac{a_{hs}^4}{ms^8} H^8$ in the effective potential, and allows for the terms of order of $H^6$ negative. Therefore, a theory with a negative $\lambda_{hs}$ may results in a negative enhancement in $\lambda_3$ as we go beyond a $(\phi^{\dagger}\phi)^3$ theory described before, as shown for instance in  the green region in Fig~\ref{fig:convergence}.

\section{measurement of the triple Higgs coupling at the LHC}
\label{sec:LHC}
The triple Higgs coupling $\lambda_3$ can be probed by the double Higgs production at the LHC. At the leading order (LO), there are two diagrams contributing to the process. The triangle diagram, which is sensitive to $\lambda_3$ and the box diagram. The two diagrams interfere with each other destructively. The LO matrix elements of the subprocess are known~\cite{Glover:1987nx,Dicus:1987ic,Plehn:1996wb}. NLO QCD corrections are known~\cite{Dawson:1998py} in an EFT approach, by applying the low energy theorem (LET)~\cite{Kniehl:1995tn} within the infinite quark mass approximation. NNLO corrections in the large quark mass limit are calculated in ~\cite{KFactor,deFlorian:2013jea,Grigo:2014jma}. Next-to-next-to-leading logarithmic (NNLL) corrections are calculated in ~\cite{deFlorian:2015moa}. For our analysis, we shall take a NNLO K-factor~=~2.27~\cite{KFactor}. 

For our analysis, we assume the double Higgs production is modified because of the altered $\lambda_3$ coupling. The double Higgs production rate could also be modified by introducing new particles that couple to gluon, and the Higgs in the loop~\cite{Dawson:2015oha,Batell:2015koa}. Those new particles change the amplitudes corresponding to the triangle diagram and the box diagram at the same time and also contribute to the single Higgs production, which is well measured at the LHC. Therefore, those contributions are constrained and tend to be small for the double Higgs production~\cite{Batell:2015koa}.

For the Higgs decays, we consider $\gamma\gamma$, $\tau^+\tau^-$, $W^+W^-$ and $b\bar{b}$ modes, which are measured in the single Higgs production at the LHC. The production rate of double Higgs is suppressed by three orders of magnitude compared to the single Higgs production at the LHC~\cite{LHCHXSWG}, so one of the two Higgs bosons needs to decay to $b\bar{b}$ for statistics, and $\gamma\gamma$, $\tau^+\tau^-$, and $W^+W^-$ modes can be considered for the other Higgs boson. We do not study the $b\bar{b}W^+W^-$ decay mode due to the overwhelming $t\bar{t}$ background, that renders a low significance ~\cite{Baglio:2012np,Dolan:2012rv}. The four b final states suffers from a large QCD background and therefore are very difficult for the LHC even in the boosted region of the Higgs, where the jet substructure techniques may be used~\cite{Dolan:2012rv}. In this work, we are therefore going to focus on the $b\bar{b}\gamma\gamma$ mode. 

The irreducible background in the $hh\rightarrow b\bar{b}\gamma\gamma$ channel include $b\bar{b}\gamma\gamma$, $t\bar{t}h(\gamma\gamma)$ and $z(b\bar{b})h(\gamma\gamma)$ processes. Considering the possibility that  a charm or light quarks fake a bottom quark, and a light jet fakes a photon, the processes $c\bar{c}\gamma\gamma$, $jj\gamma\gamma$, and $b\gamma jj$ also contribute to the background. The $t\bar{t}h$ background can be efficiently suppressed by vetoing extra jets, leptons or missing energy. Requiring the invariant mass of the two b-jets, $m_{b\bar{b}}$ and the two photons, $m_{\gamma\gamma}$ within some window of the Higgs mass helps to reduce the $Zh$ background and the QCD background. In the previous studies, a cut on the invariant mass of the two Higgs bosons, $m_{hh}$~\cite{Baglio:2012np,Barger:2013jfa,Yao:2013ika}, or some equivalent cuts were required~\cite{ATL-PHYS-PUB-2014-019} was imposed to further reject the background. In those studies, it was shown that an $O(1)$ precision in the triple Higgs boson coupling $\lambda_3$ may be achieved  at the 14~TeV
run of the LHC,  with a high integrated luminosity of order 3000~fb$^{-1}$.  

\begin{figure}[tbh]{
\includegraphics[width = 12cm, clip]{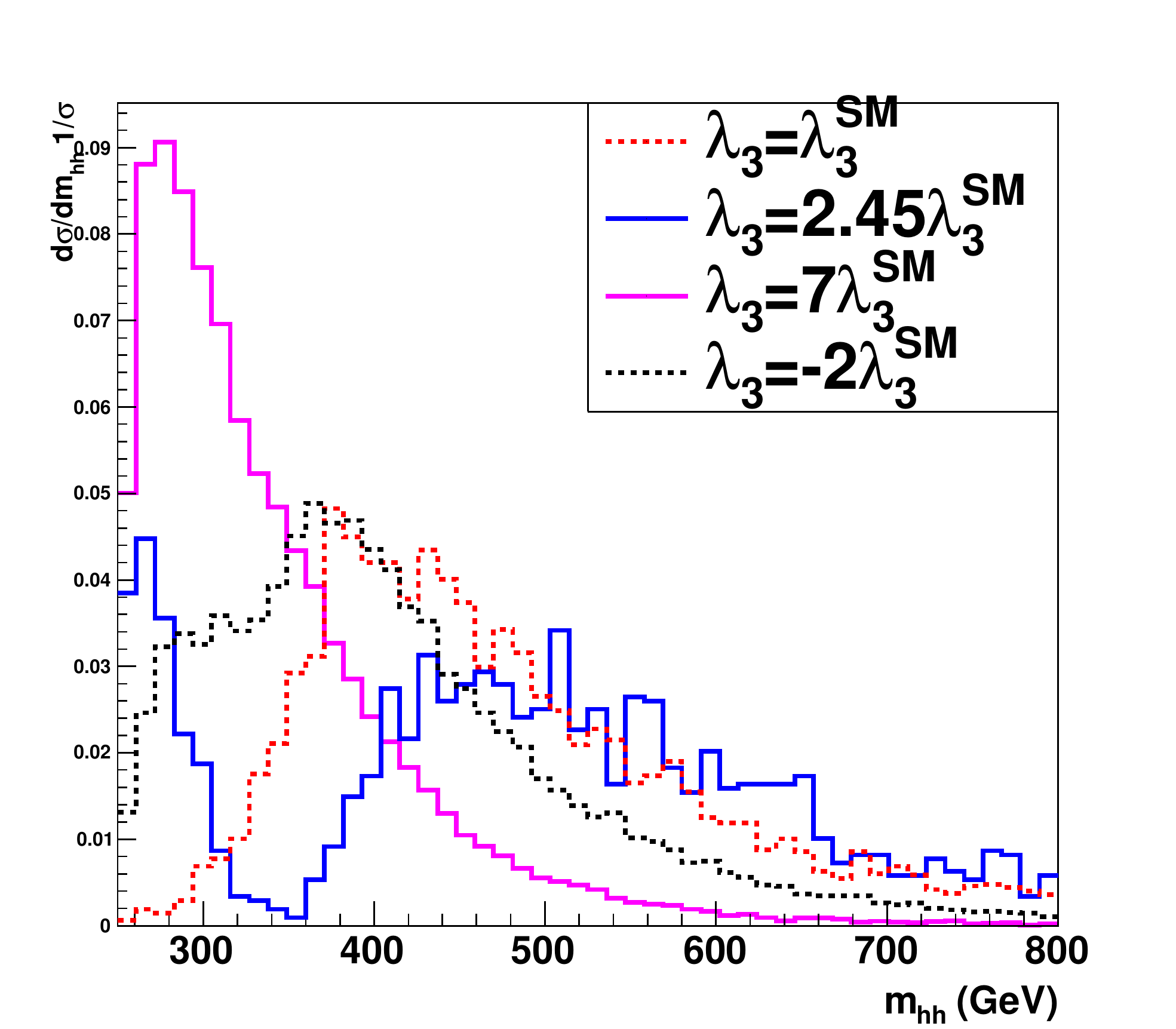}
\caption{Normalized $m_{hh}$ distributions for $\lambda_3$ = $\lambda_3^{SM}$, $\lambda_3$ = 2.45$\lambda_3^{SM}$ and $\lambda_3$ = 7$\lambda_3^{SM}$ and $\lambda_3~=~-~2~\lambda_3^{SM}$. The cancellation between the box and triangle diagram is exact at $\lambda_3$ = 2.45$\lambda_3^{SM}$ at 2$m_t$ threshold, that explains the dip. Note that the distribution shifts to smaller values as $\lambda_3$ increases}. 
\label{fig:mhh}}
\end{figure} 

As pointed out in~\cite{Barger:2013jfa}, and also noticed in ~\cite{ATL-PHYS-PUB-2014-019}, the acceptance for new physics with large $\lambda_3$ compared to the SM value is much lower for the same set of cuts. The reason for this behavior is that the $m_{hh}$ distribution is very different for the SM and for new physics with a large $\lambda_3$. When $m_{hh}$ is below the 2$m_t$ threshold, there are only real parts of the triangle and the box diagram, and these two diagrams interfere with each other destructively. The cancellation is exact at the 2~$m_t$ threshold at $\lambda_3 = 2.45 \lambda_3^{SM}$. When $m_{hh}$ is above the 2~$m_t$ threshold, imaginary parts start to develop, and the destructive interference is not as strong as it is below the 2$m_t$ threshold. So as $\lambda_3$ increases, the cross section increases more significantly below the 2~$m_t$ threshold than above the 2~$m_t$ threshold. This means that,  as $\lambda_3$ increases, the distribution of $m_{hh}$ shifts to smaller values, as shown in Fig~\ref{fig:mhh}, where we plot the normalized $m_{hh}$ distribution using MCFM~\cite{MCFM} for various values of $\lambda_3$. Thus, using the same set of cuts for new physics with a large $\lambda_3$ lead to a low acceptance at the LHC. Therefore, a modified cut on $m_{hh}$, $m_{hh} < 2m_t$ should be used when search for new physics with a large $\lambda_3$.

The $m_{hh}$ distribution also helps to distinguish positive and negative values of $\lambda_3$. For negative $\lambda_3$, the $m_{hh}$ distribution shifts to larger values  compared to the positive $\lambda_3$ that yields the same production for gluon fusion because of the constructive interference between the box and the triangle diagrams, as shown in Fig~\ref{fig:mhh}. Then, the negative and positive values of $\lambda_3$ that have the same total rate of gluon fusion can be distinguished by studying the $m_{hh}$ distribution.  


\subsection{Double Higgs production in the $b\bar{b}\gamma \gamma$ channel}

In order to understand the impact of the cuts in the  $m_{hh}$ invariant mass distribution on the reach for
double Higgs production at the LHC and future colliders, we have performed a collider study of this process
for different values of the triple Higgs coupling and in different Higgs decay channels.  In spite of the
low rate, one of the most sensitive channels is when the Higgs decays into photons, since it allows a good 
Higgs reconstruction with relatively low background.  We therefore performed 
a collider study for the $hh\rightarrow b\bar{b}\gamma\gamma$ channel. The signal with various values of $\lambda_3$ is generated by MCFM~\cite{MCFM} and passed to Pythia8~\cite{pythia} for parton shower and hadronization, and then passed to Delphes~\cite{delphes} for detector simulation. We apply a NNLO K-factor of about~2.27 for the signal~\cite{KFactor}, The background processes are generated with MadGraph~\cite{MadGraph} and then passed to Pythia and Delphes.  We apply a NLO K-factor = ~ 1.1 for $t\bar{t}h$ and a NNLO QCD, NLO EW K-factor = 1.33 for Zh~\cite{LHCHXSWG}. There are no higher order corrections known for the QCD backgrounds, and therefore, all the QCD processes are normalized to LO. We take a b-tagging efficiency of 70$\%$ and a mistag rate of 24$\%$ for c-jets and 2$\%$ for light jets~\cite{HLCMS}.  We adopt a photon tagging rate of 85$\%$  and a jet to photon fake rate $\epsilon_{j\rightarrow\gamma} = 1.2 \times 10^{-4}$~\cite{Aad:2009wy}.  We require the following cuts
\begin{eqnarray}
&p_t(b) > 30 \ \textrm{GeV}, |\eta(b)| < 2.5, \ p_t(\gamma) > 30\  \textrm{GeV} , |\eta(\gamma)| < 2.5\nonumber \\
&112.5\  \textrm{GeV} < m_{bb} < 137.5\  \textrm{GeV},\ 120 \ \textrm{GeV} < m_{\gamma\gamma} < 130\  \textrm{GeV}.
\label{eq:cut}
\end{eqnarray}
For the SM case, we further require 
\begin{equation}
m_{hh}> 350\  \textrm{GeV},
\label{eq:cutsm}
\end{equation}
while for $\lambda_3 > 3~\lambda_3^{SM}$, we require
\begin{equation}
250\  \textrm{GeV} < m_{hh} < 350\  \textrm{GeV}. 
\label{eq:cutnp}
\end{equation}
The results for LHC 14 TeV are displayed in Table~\ref{table:cutflow}. As shown in Table~\ref{table:sig}, the significance reaches 5$\sigma$ level at $\lambda_3 \sim 6.5\lambda_3^{SM}$, and $\lambda_3 \sim -0.2\lambda_3^{SM}$ at 14 TeV and 3000 fb$^{-1}$. One caveat of this analysis is that we include a K-factor for the signal (and also for the $\textrm{ZH}$ and $\textrm{tth}$ background), but the QCD background is only considered at LO. If we assume a K-factor of about 2 for the QCD processes, the significance will drop by a factor of $\sqrt{2}$, which can be compensated by the fact that there are two detectors. 
\begin{table}
\centering
\begin{tabular}{|c|c|c|c|}\hline
 & $\sigma$ (fb) &  Eq~(\ref{eq:cut}) + Eq~(\ref{eq:cutsm}) (fb)& Eq~(\ref{eq:cut}) + Eq~(\ref{eq:cutnp}) (fb)  \\
 \hline
 hh($b\bar{b}\gamma\gamma$)~($\lambda_3$ = $\lambda_3^{SM}$) &0.15 &  1.0$\times$ 10$^{-2}$ & -  \\
 hh($b\bar{b}\gamma\gamma$)~($\lambda_3$ = 5$\lambda_3^{SM}$) &0.26 & - &1.12 $\times$ 10$^{-2}$ \\
 hh($b\bar{b}\gamma\gamma$)~($\lambda_3$ = 7 $\lambda_3^{SM}$) & 0.71 & - & 3.3$\times$ 10$^{-2}$\\
 hh($b\bar{b}\gamma\gamma$)~($\lambda_3$ = 9 $\lambda_3^{SM}$) &  1.43& - & 6.08$\times$ 10$^{-2}$ \\
  hh($b\bar{b}\gamma\gamma$)~($\lambda_3$ = 0) & 0.29 & 1.33$\times$10$^{-2}$ &  -  \\ 
 hh($b\bar{b}\gamma\gamma$)~($\lambda_3 = - \lambda_3^{SM}$) & 0.50 & 2.26$\times$ 10$^{-2}$ & -  \\
  hh($b\bar{b}\gamma\gamma$)~($\lambda_3 = -2 \lambda_3^{SM}$) & 0.77 & 2.94$\times$ 10$^{-2}$ & - \\
 \hline
  $b\bar{b}\gamma\gamma$ & 5.05$\times$10$^3$ & 1.34$\times$10$^{-2}$ & 4.0$\times$10$^{-2}$ \\
 $c\bar{c}\gamma\gamma$ & 6.55$\times$ 10$^3$ & 4.19 $\times$10$^{-3} $ & 2.68$\times$10$^{-2}$\\
 $b\bar{b}\gamma j$ & 9.66$\times$10$^6$ & 4.60$\times$10$^{-3}$ & 1.38$\times$10 $^{-2}$  \\
 $jj\gamma\gamma$& 7.82$\times$10$^5$ & 2.38$\times$10$^{-3}$ & 5.26$\times$10$^{-3}$  \\
 $t\bar{t}h$ & 1.39 & 1.40$\times$10$^{-3}$ & 2.33$\times$10$^{-3}$  \\
 $zh$ & 0.33 & 6.86$\times$10$^{-4}$ & 9.01$\times$10$^{-4} $  \\
 $b\bar{b}jj$ & 7.51$\times$10$^{9}$ & 5.34$\times$10$^{-4}$ & 6.47 $\times$10$^{-4}$  \\
 \hline
\end{tabular}
\caption{Cross section in fb of the hh signal and various backgrounds expected at the LHC at $\sqrt{s}$ = 14 TeV  after applying the cuts discussed in Eq~(\ref{eq:cut}), ~(\ref{eq:cutsm}) and ~(\ref{eq:cutnp}).}
\label{table:cutflow}
\end{table}
\begin{table}
\centering
\begin{tabular}{|c|c|c|c|c|c|c|c|}\hline
$\lambda_3$&~$\lambda_3^{SM}$ & ~5$\lambda_3^{SM}$ & ~7$\lambda_3^{SM}$ &~9$\lambda_3^{SM}$ & 0 & ~-$\lambda_3^{SM}$ & ~-2$\lambda_3^{SM}$ \\
\hline
 $S/\sqrt{B}$& 3.3 & 2.1 & 6.0 & 11& 4.4 & 7.5 & 9.8\\
\hline
\end{tabular}
\caption{Significance expected for hh at the LHC at $\sqrt{s}$ = 14 TeV for an integrated luminosity of 3000~fb$^{-1}$ after applying cuts in Eq~(\ref{eq:cut}) + Eq~(\ref{eq:cutsm}) ($\lambda_3 < 3\lambda_3^{SM}$), or Eq~(\ref{eq:cut})+Eq~(\ref{eq:cutnp}) ($\lambda_3 >3~\lambda_3^{SM}$).} 
\label{table:sig}
\end{table}

It is instructive to compare these results with those obtained by the LHC experimental collaborations.
ATLAS and CMS have performed similar studies on the $hh \to bb\gamma\gamma$ channel. For HL-LHC, ATLAS expects a 1.3~$\sigma$ significance for the SM case~\cite{ATL-PHYS-PUB-2014-019}, and the CMS expectation is about 1.6~$\sigma$~\cite{CMShh}. These results are about a factor two weaker
than the ones we obtain in our study. On the other hand, 
the results from current theoretical studies show a significance range from 2$\sigma$ to 6$\sigma$~\cite{Baur:2003gp,Baglio:2012np,Barger:2013jfa,Yao:2013ika}.  The difference with the experimental results may proceed from different sources. In our analysis, we use very simple cuts, and we do not attempt to optimize the cuts for the SM background, but we believe extra cuts do not help much in this case as it is a rare process. We also do not try to perform a realistic detector simulation. 

The main issue we want to stress is the impact of the cuts in the invariant mass distribution when studying possible modifications of the triple 
Higgs coupling.  We obtain a very significant sensitivity improvement in the case where $\lambda_3$ deviates significantly from the SM, when we implement the new cuts in Eq.~(\ref{eq:cutnp}) we propose for such cases. For instance, when $\lambda_3 = 5\lambda_3^{SM}$, if we use the cuts in Eq.~(\ref{eq:cutsm}), we only expect a 0.67$\sigma$ significance, while we expect 2.1$\sigma$  significance if we use the cuts in Eq.~(\ref{eq:cutnp}).   Similar large improvements are obtained for other sizable values of $\lambda_3 > 3 \lambda_3^{SM}$. 

Due to the relatively low sensitivity of the LHC in looking for double Higgs production, it is interesting to consider similar signatures at future colliders, in particular a future high energy $pp$ collider.  The sensitivity will depend on many factors, including the center of mass energy and the detector performance.  To be specific, we shall consider the case of  100~TeV  $pp$ collider, assuming that the detector performance stays the same as at the LHC, performing similar cuts as the ones in the LHC analysis. We show the results in Table~\ref{table:FCC} and Table~\ref{table:FCCsig}. In our analysis, we considered only positive values of $\lambda_3$, since as shown above, the LHC is already sensitive to the negative values. It is then easy to extrapolate the same analysis for higher energies. The results presented in Table~\ref{table:FCC} show that a 100~TeV collider should be sensitive to triple Higgs boson couplings $\lambda_3 \sim 5 \lambda_3^{SM}$,  where the  same cuts proposed in Eq~(\ref{eq:cut}) were used.   The significance we
obtain is similar the ones obtained in Refs.~\cite{Barr:2014sga} and~\cite{He:2015spf} for
the same process. Again, we obtain a significant improvement of the sensitivity at large values of $\lambda_3 > 3 \lambda_3^{SM}$ when the new cuts on $m_{hh}$ given in Eq.~(\ref{eq:cutnp}) are used. 

\begin{table}
\centering
\begin{tabular}{|c|c|c|c|}\hline
 & $\sigma$  (fb) &  Eq~(\ref{eq:cut}) + Eq~(\ref{eq:cutsm}) (fb) & Eq~(\ref{eq:cut}) + Eq~(\ref{eq:cutnp}) (fb) \\
 \hline
 hh($\lambda_3$ = $\lambda_3^{SM}$) &3.4 & 0.11 & -  \\
 hh($\lambda_3$ = 3$\lambda_3^{SM}$) & 1.48 & 0.042 & - \\ 
 hh($\lambda_3$ = 5$\lambda_3^{SM}$) &4.45 & - &0.10 \\
 \hline
 $b\bar{b}\gamma\gamma$ & 1.7$\times$10$^6$ & 0.129 & 0.52 \\
 $c\bar{c}\gamma\gamma$ &1.0$\times$10$^5$ & 6.45 $\times$10$^{-2} $ &0.42  \\
 $b\bar{b}\gamma j$ & 1.19$\times$10$^5$ & 1.68$\times$10$^{-2}$ & 6.72$\times$10$^{-2}$  \\
 $jj\gamma\gamma$& 2.73$\times$10$^6$ &1.92$\times$10$^{-2}$ & 7.3$\times$10$^{-2}$\\
 $t\bar{t}h$ & 86.41 & 2.72$\times$10$^{-2}$ & 2.53$\times$10$^{-2}$ \\
 $zh$ &0.88 & 1.76$\times$10$^{-3}$ & 1.4$\times$10$^{-3} $  \\
 $b\bar{b}jj$ & 4.07$\times$10$^{10}$ & 2$\times$10$^{-3}$ & 4.7 $\times$10$^{-3}$  \\
 \hline
\end{tabular}
\caption{Cross section of the hh signal and various backgrounds expected at a 100 TeV collider after applying the cuts discussed in Eq~(\ref{eq:cut}), ~(\ref{eq:cutsm}) and ~(\ref{eq:cutnp}). }
\label{table:FCC}
\end{table}

\begin{table}
\begin{tabular}{|c|c|c|c|}\hline
$\lambda_3$ &$ ~\lambda_3^{SM}$ & ~3$\lambda_3^{SM}$  &~5$\lambda_3^{SM}$ \\
\hline
$S/\sqrt{B}$ & 11 & 4.5 & 5.3\\
\hline
\end{tabular}
\caption{The significance of double Higgs production expected for hh at a 100 TeV collider for an integrated luminosity of $3000\,fb^{-1}$ after applying cuts in Eq~(\ref{eq:cut}) + Eq~(\ref{eq:cutsm}) ($\lambda_3<3\lambda_3^{SM}$), or Eq~(\ref{eq:cut}) + Eq~(\ref{eq:cutnp}) ($\lambda_3 >3~\lambda_3^{SM}$)}.
\label{table:FCCsig}
\end{table}

\subsection{Double Higgs production in the $b\bar{b}\tau^+\tau^-$ channel}

Since the Higgs has many different significant decay channels, it is useful to think about double Higgs production in channels different from the
$bb \gamma \gamma$ considered in this work.  A particularly interesting one is the $bb\tau\tau$ channel.  The $b\bar{b}\tau^+\tau^-$ channel enjoys a larger cross section but suffers from the difficulty in the event reconstruction due to the missing energy associated with $\tau$ decays.  It also suffers from
larger backgrounds that should be properly considered to obtain a realistic reach estimate.

 The $\tau$ pair invariant mass $m_{\tau\tau}$ may  be estimated by the missing mass calculator~\cite{MMC}, and similar methods could be used to estimate $m_{hh}$ in this channel. In order to estimate the reach in this channel, we shall assume that the $m_{\tau\tau}$  invariant mass can be reconstructed with a similar resolution as $m_{bb}$~\cite{MMC} invariant mass. Furthermore, we shall assume that the two Higgs invariant mass  $m_{hh}$ can be reconstructed as well as it is obtained at the parton level.  The discovery reach is then  estimated adopting the cuts and background calculations presented
 in Ref.~\cite{Baglio:2012np}.
 
We go beyond the analysis of Ref.~\cite{Baglio:2012np} by including the relevant background coming from the $bbjj$ process. Under
 the above conditions, and assuming a jet to $\tau$ fake rate $\epsilon_{j\rightarrow\tau} = 1/100$~\cite{Barger:2013jfa}, we obtain a significance $S/\sqrt{B} \sim 3.75$ for $\lambda_3 = \lambda_3^{SM}$, that is similar to the one obtained in the $\gamma \gamma$ channel. However, estimating $m_{hh}$ in the $bb\tau\tau$ channel is very difficult. For that reason, CMS preforms a preliminary study using the Stransverse mass $m_{T2}$ instead of $m_{hh}$ to distinguish the signal from the background, and shows a 0.9$\sigma$ significance for HL-LHC~\cite{CMShh}. That is significantly smaller than the one found in~\cite{Barr:2013tda} using a similar method. Therefore, the $bb \tau\tau$ channel may represent a good complementary channel to the $bb\gamma\gamma$ one, and should be studied further.  
\section{Conclusions}
\label{conclusions}

In this work, we have studied the modifications of the triple Higgs couplings in theories in which the Higgs potential is modified by the addition of higher order, non-renormalizable operators, induced by the presence of new physics at the weak scale. Contrary to previous statements in the literature, we have shown that, a simple addition of a dimension six operators may lead to a large modification of the triple Higgs coupling $\lambda_3$ with respect to its SM value in the regions of parameter space consistent with a FOEPT. 

Furthermore, the addition of higher order operators may also lead to a reduction of the triple Higgs coupling, or even its change of sign, with relevant implications for collider physics. Interestingly, negative enhancements of the triple Higgs coupling tend to be associated with a second order phase transition, while a first order
phase transition tends to be associated with a large positive enhancement of this coupling. 

We also argue, building up on the previous results in the literature, that different values of the triple Higgs coupling will have a strong impact not only on the total cross section, but also on the invariant mass distribution of double Higgs production at the LHC. This motivates the use of different cuts for double Higgs production for values of the trilinear coupling about or smaller than the SM value than for the large values of $\lambda_3$. The determination of the total cross section, together with the analysis of the invariant mass distribution may give hints not only about the magnitude of the departure of the Higgs coupling with respect to the SM value, but also of its sign. Considering these different cuts in the invariant mass distribution and including background processes that were previously ignored in the literature, we showed that at the 14~TeV run of the LHC at high luminosities of order of a 3.3$\sigma$ is expected for $\lambda_3 = \lambda_3^{SM}$, and a 5 $\sigma$ significance is expected for $\lambda_3 = 6.5\lambda_3^{SM}$ (-$0.2~\lambda_3^{SM}$) for the $b\bar{b}\gamma\gamma$ channel. The $b\bar{b}\tau^+\tau^-$ channel presents a promising complementary channel.

\section{acknowledgments}
We would like to thank V. Barger, L. Everett, J. Gao, H. Haber, A. Ismail, I. Lewis, M.~Peskin and L.T. Wang for useful discussions. Work is supported by the U.S. Department of Energy under Contract No. DE-FG02-13ER41958. Work at ANL is supported in part by the U.S. Department of Energy under Contract No. DE-AC02-06CH11357.  P.H. is partially supported by U.S. Department of Energy Grant DE-FG02-04ER41286.

\newpage
\appendix
\allowdisplaybreaks
\section*{Appendix}
\section{Triple Higgs Coupling}\label{app:A}
We add tree-level non renormalizable operators to the Higgs potential to get the most general effective potential at the tree-level
\begin{align}
V\left(\phi\right)=&\sum^\infty_{n=1}\frac{k_{2n}}{2n}\phi^{2n},\label{eqn:tadpole}
\end{align}
where $k_2 = m^2$, $k_4 = \lambda$ and, for $n \geq 3$,
\begin{equation}
\frac{k_{2n}}{2n} = \frac{c_{2n}}{2^n \Lambda^{2(n-2)}}\label{eqn:ktoc}
\end{equation}
For the potential to have minimum at the VEV it must satisfy
\begin{align}\label{eqn:tadpole1}
\frac{\partial V}{\partial\phi}\bigg|_{\phi=v}=\sum^\infty_{n=1}k_{2n}v^{2n-1}=0&.
\end{align}

The second derivative at the VEV must be the square of the Higgs boson mass as discovered by the CMS and ATLAS experiments at the LHC~\cite{Chatrchyan:2012xdj,Aad:2012tfa}
\begin{align}
\frac{\partial^2 V}{\partial\phi^2}=&\sum^\infty_{n=1}(2n-1)k_{2n}\phi^{2n-2},&\notag\\
\frac{\partial^2 V}{\partial\phi^2}\bigg|_{\phi=v}=&\sum^\infty_{n=1}(2n-1)k_{2n}v^{2n-2}=m^2_h.\label{eqn:mass1}
\end{align}
Dividing~\ref{eqn:tadpole1} by $v$ and then subtracting it from~\ref{eqn:mass1}, we get
\begin{align}
\sum^\infty_{n=2}(2n-2)k_{2n}v^{2n-2}=m^2_h,\notag\\
\sum^\infty_{n=2}(n-1)k_{2n}v^{2n-4}=\frac{m^2_h}{2v^2}.\label{eqn:k4}
\end{align}
The third derivative will give the triple Higgs coupling as we are already in the canonical normalization, where we can substitute $\phi=h+v$ and $v=246$ GeV. 
\begin{align}\label{eqn:triple1}
\frac{\partial^3 V}{\partial\phi^3}\bigg|_{\phi=v}=&\sum^\infty_{n=2}(2n-1)(2n-2)k_{2n}v^{2n-3}.
\end{align}
Multiplying~\ref{eqn:k4} by $6v$ and subtracting it from~\ref{eqn:triple1} we get
\begin{align}
\lambda_3=&\,\,\,\frac{\partial^3 V}{\partial\phi^3}\bigg|_{\phi=v}=\frac{3m^2_h}{v}\left(1+\sum^\infty_{n=3}\frac{4(n-1)(n-2)k_{2n}v^{2(n-1)}}{3m^2_h}\right).\label{eqn:triplek}
\end{align}
Substituting for $k$ in terms of the cut-off of the effective theory $(\Lambda$) and the corresponding dimensionless coefficients ($c_{2n}$) from Eq.~(\ref{eqn:ktoc}), we obtain
\begin{align}
\lambda_3=&\,\,\frac{3m^2_h}{v}\left(1+\frac{8v^2}{3m_h^2}\sum^\infty_{n=3}\frac{n(n-1)(n-2)c_{2n}v^{2(n-2)}}{2^n\Lambda^{2(n-2)}}\right),\label{eqn:triple2}
\end{align}
where $|c_{2n}|<1$. This can be written as
\begin{align}
\lambda_3&=\frac{3m^2_h}{v}\left(1+\frac{8\Lambda^2}{3m_h^2v^2}\sum^\infty_{n=3}n(n-1)(n-2)c_{2n}\left(\frac{v^2}{2\Lambda^2}\right)^n\right).\label{eqn:triple3}
\end{align}
From this we clearly see that the the series converges, even if all $c_{2n}$ are 1, for 
\begin{align}
\Lambda>\frac{v}{\sqrt{2}}\sim174\,\text{GeV}.
\end{align}

\section{Maximal Negative Enhancements of $\lambda_3$ for $(\phi^\dagger\phi)^4$ and $(\phi^\dagger\phi)^5$}\label{app:C}

The value of the triple Higgs coupling $\lambda_3$ is associated with the third derivative of the potential at the minimum, which corresponds to the change in the potential curvature. At the minimum of the Higgs potential at the VEV, the curvature value is a measured positive constant. Therefore, a negative $\lambda_3$ implies even lower curvatures for the higher values of $\phi$. In the extreme case, where the curvature turns negative,  this will generate a maximum. Hence there has to be one more minimum for even higher values of $\phi$ sot that the potential is stable in the limit of $\phi\rightarrow\infty$. Let the position of such a minimum be $\phi=p$. 

This potential can be written as
\begin{align}
v(\phi)&=\frac{k_8}{8}\left(\phi^2-v^2\right)^2\left(\phi^2-p^2\right)^2-\frac{k_8}{8}v^4p^4
\end{align}
Comparing this expression  with the generic form of the Higgs potential,  Eq.~(\ref{eqn:pot8}), we get
\begin{align}
\frac{k_6}{6}=-\frac{3k_8}{4}(p^2+v^2),\quad\frac{\lambda}{4}=-\frac{k_8}{8}(p^4+v^4+4p^2v^2)\label{eqn:k6k4}
\end{align}
Substituting this in Eq.~(\ref{eqn:k4}) we obtain a relation between $k_8$ and the Higgs mass, namely
\begin{align}
k_8=\frac{m_h^2}{v^2(p^2-v^2)^2}
\end{align}
Substituting in Eq.~(\ref{eqn:k6k4}) gives
\begin{align}
k_6=-\frac{3m_h^2}{2v^2}\frac{(p^2+v^2)}{(p^2-v^2)^2}\label{eqn:k6p}
\end{align}
$k_8$ has to be positive for the stability of the potential. Therefore $k_6$ is the only term that contributes to the enhancement with opposite sign. The maximal negative value it can take is, for $c_6 < 1$,
\begin{align}
k_6=-\frac{3}{4\Lambda^2},\label{eqn:k6lamb}
\end{align}
Equating the right hand sides of the Eqs.~(\ref{eqn:k6p}) and Eq.~(\ref{eqn:k6lamb}) yields 
\begin{align}
2m_h^2(p^2+v^2)\Lambda^2=(p^2-v^2)^2v^2
\end{align}
Solving for $p^2$ gives
\begin{align}
p^2-v^2=\frac{m_h^2\Lambda^2\pm\sqrt{m_h^4\Lambda^4+4m_h^2\Lambda^2v^4}}{v^2}
\end{align} 
The right hand side must be greater than 0 as $p>v$. This implies
\begin{align}
p^2-v^2=\frac{m_h\Lambda}{v^2}\left(m_h\Lambda+\sqrt{m_h^2\Lambda^2+4v^4}\right)\label{eqn:p-v}
\end{align}
From Eq.~(\ref{eqn:triplek}), we know
\begin{align}
\frac{\lambda_3}{\lambda_3^{SM}}-1=\frac{8v^4}{3m_h^2}(k_6+3k_8v^2)\label{eqn:correctionphi8}
\end{align}
Substituting Eq.~(\ref{eqn:k6p}) in Eq.~(\ref{eqn:correctionphi8}) gives
\begin{align}
\frac{\lambda_3}{\lambda_3^{SM}}-1=-\frac{4v^2}{p^2-v^2}
\end{align}
Using Eq.~(\ref{eqn:p-v}), we get the maximum negative enhancement, namely
\begin{align}
\delta=\frac{\lambda_3}{\lambda_3^{SM}}-1=-\frac{x}{1+\sqrt{1+x}},\quad\text{where}\quad x=\frac{4v^4}{m_h^2\Lambda^2}.
\end{align}

\bibliographystyle{utphys}
\bibliography{HHref}
\end{document}